\documentclass[a4paper,12pt,oneside]{article}

\usepackage{amsmath, bm}
\usepackage{amsfonts}
\usepackage[mathscr]{euscript}
\usepackage[T1]{fontenc}
\usepackage[pdftex]{graphicx}
\usepackage{pdfpages}
\usepackage{graphicx}
\usepackage{cancel}
\usepackage{xcolor}
\usepackage[top=25mm,right=25mm,left=25mm,bottom=25mm]{geometry}
\usepackage{caption}
\usepackage{subcaption}
\usepackage{setspace}
\usepackage[pdftex]{hyperref}
\usepackage[affil-it,auth-sc]{authblk}
\usepackage{gensymb}

\usepackage[autostyle]{csquotes}
\usepackage[
    backend=bibtex,
    style=numeric,
    citestyle=numeric-comp,
    maxbibnames=99,
    minbibnames=99,
    maxcitenames=2,
    mincitenames=1,
    date=year,
    giveninits=true,
    sorting=none,
    sortlocale=auto,
    natbib=true,
    url=false, 
    isbn=false,
    doi=true,
    eprint=true
]{biblatex}
\DeclareMathOperator{\tr}{tr}
\DeclareMathOperator{\dev}{dev}

\graphicspath{{./}{./figures/}}                         

\title{Thermomechanical modelling of ceramic pressing and subsequent sintering}

\author[]{D. Kempen}
\author[]{A. Piccolroaz}
\author[]{D. Bigoni\footnote{Corresponding author: e-mail: \href{mailto:bigoni@ing.unitn.it}{bigoni@ing.unitn.it}; phone: +39\,0461\,282507.}}

\affil[]{Department of Civil, Environmental and Mechanical Engineering, University of Trento, Italy}

\date{}

\begin{document}

\maketitle


\begin{abstract}
An elastic-visco-plastic thermomechanical model for the simulation of cold forming and subsequent sintering of ceramic powders is introduced and based on micromechanical modelling 
of the compaction process of granulates. Micromechanics leads to an upper-bound estimate of the compaction curve of a granular material, which 
compares well with other models and finite element simulations. The parameters of the thermomechanical model are determined on the basis of available data and 
dilatometer experiments. Finally, after computer implementation, validation of the model is performed against experiments developed on specially designed ceramic pieces, characterized by zones of different density. 
The mechanical model is found to accurately describe forming and sintering of stoneware ceramics and can therefore be used to analyze and optimize 
industrial processes involving compaction of powders and subsequent firing of the greens.
\end{abstract}

\section{Introduction}

The production of ceramic pieces is based on technologies involving a massive waste of energy and materials\footnote{
Grinding of the raw material requires mills with a power up to 1 MW, while drying and increasing of the temperature of the slip involves up to 500 KW of
electrical power and 15 Gcal/h of thermal power. The 80$\%$ of the thermal capacity is lost at the chimney and powder is spread in the environment. The
forming of the ceramic powders represents a huge waste of energy, because only the 5$\%$ of the energy is transmitted to the final piece from the presses (up to 250 KW
of installed power). Finally, drying and sintering requires large burners consuming up to 10 Gcal/h.
}, 
so that environmental preservation 
imposes a rationalization of industrial processes to reduce pollution. 
The optimization of the production process is directly linked to the availability of models for the mechanical behaviour of the powders and binders used during
forming, including the simulation of both phases of cold compaction and subsequent sintering, and in the design of mechanical characteristics of the final pieces. 

Sintering is the common method for completing the process of ceramic
production and involves heating of the green bodies, obtained in a preceding compaction stage, to obtain the
required density and strength of the final products. Understanding and modelling the mechanisms involved in the processes of compaction and 
sintering is therefore crucial to ensure high quality and reproducibility of ceramic materials. 

In the last 50 years, there have been significant developments in the theory of sintering, but only in the 1980s the use of
continuum mechanics was introduced, to 
predict the stresses and strains that can develop during the process. A comprehensive review of the models
for sintering proposed in the last 25 years can be found in \cite{Guillon2007}. 
In recent years, meso scale models have been developed using multi particle Finite Element (MPFEM) approaches \cite{Procopio2005} or discrete element approaches \cite{Wonisch2009}. These models are useful in the understanding of micro- or meso-mechanical effects, but, due to the limitation in the number of particles that can analyzed, they are hardly applicable to model whole complex shaped parts, so that continuum models, such those developed in \cite{Abouaf1988,Duva1992,Olevsky1998}, play a central role. 

Often sintering was studied with reference to isothermal conditions \cite{Olevsky1998,Kraft2004}, but during the firing the body is subject to non-isothermal loadings, especially for high heating rates and large ceramic pieces, so that the modelling of this situation is one of the objectives of the present article.

Before sintering, a green body is obtained through powder compaction and, depending on the geometry and the compaction method, the pressed green body shows usually significant differences of local density throughout the part 
\cite{Stupkiewicz2015,Scot_Swan_2017}, so that powder pressing 
has a strong influence on sintering, which depends on the (relative) density of the green. 
Therefore a continuum model that can predict the density distribution after the pressing step and the subsequent sintering is important for industrial applications \cite{Penasa_2016}.
In this research direction, while cold compaction has been often addressed \cite{Fleck1995,Barbara1992,Redanz2001,Lewis2005,Piccolroaz_2006,Piccolroaz_2006_2,Cocks2007,salvadori2017numerical,Krairi_2018}, the combination of pressing and sintering 
has been scarcely investigated. 
In particular, two different models, one for powder pressing and another for sintering have been proposed 
\cite{Kraft2004}, while 
thermomechanical models for hot-isostatic pressing of metal powders were developed using a pressure-sensitive and temperature dependent yield function \cite{Mahler2003,Frischkorn2008}.

The objective of the present article is to propose a thermomechanical, elastic-visco-plastic model to simulate cold powder compaction and subsequent non-isothermal solid state sintering. 
The model is grounded on the thermomechanical framework developed by Simo and Miehe \cite{Simo1992, Miehe2002} and on the additive decomposition of elastic and plastic logarithmic strains \cite{Miehe2002, Sansour2003}. Moreover, the BP yield function \cite{Bigoni2004} is used to simulate both compaction and firing, so that these two process steps become more closely integrated. With reference to an aluminum silicate spray dried powder (used for the industrial production of ceramic tiles), 
the model is calibrated against both available data and {\it ad-hoc} performed experiments and is implemented (through an UMAT routine) in a finite element code (Abaqus). 
Finally, validation against experiments, conducted on large ceramic pieces (in which density variations were intentionally introduced), shows  
that the model provides an accurate prediction of the entire process of forming and sintering and therefore remains now 
available for the design and optimization of ceramic pieces.

\section{Thermoinelastic model for the sintering of ceramics}

With the purpose of introducing a model for sintering of green ceramics, the large deformation of a thermo-elastic-visco-plastic solid is considered, described by the deformation gradient $\bm{F}$, the right Cauchy-Green tensor and the corresponding Lagrangean logarithmic strain
\begin{equation}
\bm{C} = \bm{F}^T\bm{F}, ~~~~ \bm{\epsilon} = \frac{1}{2} \log \bm{C},
\end{equation}
where the superscript $T$ denotes the transpose.

A key assumption is the additive decomposition of logarithmic strain  into an elastic (subscript \lq $e$') and a visco-plastic (subscript \lq $p$') component proposed by Miehe et al. \cite{Miehe2002} and Sansour and Wagner \cite{Sansour2003} as
\begin{equation}
 \label{eq:SmallStrain}
{\bm{\epsilon}}={\bm{\epsilon}}_{e}+{\bm{\epsilon}}_{p}. 
\end{equation}

The Helmholtz free-energy $\psi(\bm{\epsilon}_e, T,\widehat{\rho})$ is introduced as a function of (i.) the elastic part of the logarithmic strain, (ii.) the temperature $T$, 
and (iii.) the inelastic relative density of the material $\widehat{\rho}$, which is the only internal variable introduced in the treatment, 
defined as 
\begin{equation}
\label{eq:DefinitionRhoRel}
\widehat{\rho} =\frac{\rho_0}{\rho_{fd}} e^{-\tr \bm{\epsilon}_p},
\end{equation}
and representing a dimensionless measure of the mass density upon unloading. In Eq. (\ref{eq:DefinitionRhoRel}) $\rho_0$ is the initial value of the mass density $\rho$, and $\rho_{fd}$ the value corresponding to the fully dense material, which contains no pores.

Definition (\ref{eq:DefinitionRhoRel}) may be better appreciated by writing the rate of mass conservation
\begin{equation}
\frac{\dot{\rho}}{\rho} = -\tr \dot{\bm{\epsilon}},
\end{equation}
which can be integrated in time to provide the  expression
\begin{equation}
\label{zio}
\log \frac{\rho}{\rho_0} = -\tr \bm{\epsilon}. 
\end{equation}
From eq. (\ref{zio}) the rate of mass density can be calculated to be 
\begin{equation}
\frac{\dot{\rho}}{\rho_{fd}} =\left[-\widehat{\rho} \tr \dot{\bm{\epsilon}}_e + (\widehat{\rho})^\cdot \right]e^{-\tr \bm{\epsilon}_e} ,
\end{equation}
where
\begin{equation}
(\widehat{\rho})^\cdot =- \widehat{\rho} \tr \dot{\bm{\epsilon}}_p,
\end{equation}
an equation which will become useful later. 

Note also that the mass density is related to the porosity $f$ (the ratio between the volume of the voids and the total volume of a sample) of the material through the equation
\begin{equation}
\label{effe}
\frac{\rho}{\rho_{fd}} = 1 - f.
\end{equation}
Eq. (\ref{effe}) can be understood considering the deformation gradient $\bm{F}_{fd}$ and its determinant $J_{fd}$ needed to bring the current volume element $V$ to the volume of the fully dense material $V_{fd}$. In this deformation, the volume and the density transform according to the well-known rules $V_{fd} = J_{fd}V$ and $\rho = J_{fd} \rho_{fd}$, respectively, so that eq. (\ref{effe}) is obtained, because by definition $f=1-V_{fd}/V$.

The so-called elastoplastic coupling (in which the plastic strain influences the elastic stiffness, \cite{Stupkiewicz_2014,Gajo2015,Argani_2016}) is not introduced in the model and the elastic part of the deformation (not particularly important during sintering) will be eventually treated with the standard linear isotropic thermoelastic law \cite{Doghri2000}. 
Therefore, the stress $\bm{\sigma}$, work-conjugate in the Hill sense to the 
Lagrangian logarithmic strain, the thermodynamical force $R$ associated to the 
internal variable $\widehat{\rho}$, and the entropy $\eta$ can be expressed as
\begin{equation}
\label{sigmone}
\bm{\sigma} = \rho \frac{\partial \psi(\bm{\epsilon}_e, T,\widehat{\rho})}{\partial \bm{\epsilon}_e}, ~~ R = -\rho \frac{\partial \psi(\bm{\epsilon}_e, T,\widehat{\rho})}{\partial \widehat{\rho}}, ~~ \eta = -\frac{\partial \psi(\bm{\epsilon}_e, T,\widehat{\rho})}{\partial T}.
\end{equation}

For simplicity, an additive form of the Helmholtz free energy is assumed, sum of the elastic and the purely thermal energies, plus a \lq pore energy' 
\begin{equation}
\label{eq:FreeEnergyAsFunctionOfStateVariables}
 \psi(\bm{\epsilon}_e, T, \widehat{\rho}) = \psi_{e}(\bm{\epsilon_{e}},T)+\psi_{T}(T)+\psi_{pore}(\widehat{\rho}),
\end{equation}
where
\begin{equation}
\psi_{T}(T) = - c_h T \log\frac{T}{T_0}+c_h (T-T_0) - \eta_0 (T-T_0) + \psi_0,
\end{equation}
in which $T_0$ is the absolute temperature corresponding to the unstressed material when $\bm{\epsilon}=0$; moreover, $\eta_0$ and $\psi_0$ are the values of entropy $\eta$ and free energy at $T=T_0$ and $\bm{\epsilon}=0$; finally 
\begin{equation}
\label{ci}
c_h = -T \frac{\partial^2 \psi(\bm{\epsilon}_e, T,\widehat{\rho})}{\partial T^2},
\end{equation}
is the specific heat at constant values of strain and internal variables. 

The density, together with the (visco-)plastic strain $\bm{\epsilon}_{p}$  are internal variables of the system. The external variables are the total strain $\bm{\epsilon}$ and the temperature $T$.
To follow the theory of thermodynamics of irreversible processes, the Clausius-Duhem dissipation inequality has to be fulfilled at all times \cite{Maugin1992,Doghri2000}, namely
\begin{equation} 
 \label{eq:Dissipationinequality}
 - \rho \left( \dot{\psi} + \eta \dot{T} \right) + \bm{\sigma} \cdot \dot{\bm{\epsilon}} - \frac{1}{T} \, \bm{q} \cdot \nabla T \geq 0.
\end{equation}
The rate in time of the Helmholtz free energy $\psi$, eq. (\ref{eq:FreeEnergyAsFunctionOfStateVariables}), can be expressed in terms of its state variables:
\begin{equation}
 \label{eq:FreeEnergyDerivative}
 \dot{\psi} = \frac{\partial \psi_{e}}{\partial \bm{\epsilon}_{e}}\dot{\bm{\epsilon}}_{e}
 + \frac{\partial \psi_T}{\partial T}\dot{T}  + \frac{\partial \psi_{pore}}{\partial \widehat{\rho}}(\widehat{\rho})^\cdot .
\end{equation}

The additive decomposition of the strain into an elastic and a (visco-)plastic part, eq.(\ref{eq:SmallStrain}),
can be inserted into eq.(\ref{eq:FreeEnergyDerivative}) and the result substituted into eq.(\ref{eq:Dissipationinequality}) to yield
\begin{equation}
 \label{eq:FreeEnergySplitStrainRate3}
- \underbrace{\left( \rho \frac{\partial \psi_T}{\partial T} +\rho \eta \right) \dot{T}  -\frac{1}{T} \, \bm{q} \cdot \nabla T}_{\mathrm{thermal \  dissipation}} + 
\underbrace{\left( - \rho \frac{\partial \psi_e}{\partial \bm{\epsilon}_{e}} +  \bm{\sigma} \right) \cdot \dot{\bm{\epsilon}}_{e}}_{\mathrm{elastic \ dissipation}}
 + \underbrace{ \left( {\rho \, \widehat{\rho}}\frac{\partial \psi_{pore}} {\partial \widehat{\rho}} \bm{I}+  \bm{\sigma} \right)\cdot\dot{\bm{\epsilon}}_{p} }_{\mathrm{(visco-)plastic \ dissipation}}\geq 0.
\end{equation}

Setting the thermal dissipation to be independent of $\dot{T}$ yields equation 
(\ref{sigmone})$_3$, while imposing the vanishing of 
the elastic dissipation for every $\dot{\bm{\epsilon}}_{e}$ in equation (\ref{eq:FreeEnergySplitStrainRate3}) provides equation 
(\ref{sigmone})$_1$, so that 
using the following simple expression for the potential (borrowed from the linear theory) of the elastic part of the free energy density,
\begin{equation}
\label{eq:ThermoelasticStress}
\rho  \psi_{e} =  \frac{1}{2} \lambda \left( \tr {\bm{\epsilon_{e}}} \right)^2 + \mu \bm{\epsilon}_{e} \cdot \bm{\epsilon}_{e} - K_b \, \alpha_0 \left(T-T_0 \right) \tr \bm{\epsilon}_{e},
\end{equation}
the stress is given by  the usual isotropic thermoelastic relation
\begin{equation}
\label{eq:ThermoElasticStress3}
 \bm{\sigma} =  \mathbb{C}[\bm{\epsilon}_{e}] - K_b \alpha_0 \left( T - T_0 \right) \bm{I},
\end{equation}
where $K_b$ is the elastic bulk modulus, $\alpha_0$ is the thermal expansion 
coefficient, $T_0$ a reference temperature, and the fourth-order elastic tensor 
$\mathbb{C}$ is
\begin{equation}
\label{eq:ThermoElasticStress2}
\mathbb{C} =   \lambda \, \bm{I} \otimes \bm{I} + 2 \mu \mathbb{S},
\end{equation}
in which $\mathbb{S}$ is the fourth-order symmetrizer and $\lambda$ and $\mu$ are the Lam\'e elastic moduli. 

Equation \ref{eq:ThermoElasticStress2} represents a strong assumption, which is justified in the present context, because the elastic strain and rotation are usually small during thermoplastic pressing and sintering of ceramics.

As a conclusion, the dissipation inequality eq. (\ref{eq:FreeEnergySplitStrainRate3}) reduces to 
\begin{equation}
 \label{oca}
-\frac{1}{T} \, \bm{q} \cdot \nabla T+ \left( {\rho \, \widehat{\rho}}\frac{\partial \psi_{pore}} {\partial \widehat{\rho}} \bm{I}+  \bm{\sigma} \right)\cdot\dot{\bm{\epsilon}}_{p} \geq 0.
\end{equation}

\subsection{Effective stress for sintering and dissipation}

An inspection of eq. (\ref{oca}) and consideration of eq. (\ref{sigmone})$_2$ reveals that the thermodynamic dual force to the plastic strain rate is not the stress, but an effective stress defined as \cite{Mahler2003,Frischkorn2008}
\begin{equation}
\label{eq:EffectiveStress}
 \hat{\bm{\sigma}}=  \bm{\sigma} - \sigma_s\bm{I},
\end{equation}
where 
\begin{equation}
\sigma_s = \widehat{\rho}R = -\widehat{\rho} \rho \frac{\partial \psi}{\partial 
\widehat{\rho}}
\end{equation}
is the so-called \lq sintering stress' (also known as \lq Laplace pressure' \cite{galuppi2014combined}). Eventually, the dissipation inequality, eq. (\ref{oca}), can be rewritten as 
\begin{equation}
 \label{morta}
-\frac{1}{T} \, \bm{q} \cdot \nabla T+ \hat{\bm{\sigma}}\cdot\dot{\bm{\epsilon}}_{p} \geq 0,
\end{equation}
which highlights the fact that the inequality is always a-priori satisfied, when the Fourier law of heat conduction is assumed
\begin{equation} 
\label{fou}
\bm{q} = - k \nabla T,
\end{equation}
(in which $k>0$ is the thermal conductivity), together with the normality rule for $\dot{\bm{\epsilon}}_p$ and convexity of the yield function, both 
in the $\hat{\bm{\sigma}}$-space.

\subsection{Helmholtz free energy for porosity variation}

Following \cite{Mahler2003}, the Helmholtz free energy related to the porosity, $\psi_{pore}$, can be assumed to be a linear function of the surface tension $\gamma_s$ relative to the area of the pore $A_{pore}(\widehat{\rho})$ as follows 
\begin{equation}
 \label{eq:SurfacePotential}
 \psi_{pore}=\frac{\gamma_s A_{pore}(\widehat{\rho})}{\rho V}=\frac{\gamma_s A_{pore}(\widehat{\rho})}{\rho_{fd} V_{solid}} .
\end{equation}
Assuming spherical pores of radius $r$ in a cubic unit cell, it follows that $A_{pore} =4 \pi r^2$ and assuming incompressibility of the solid phase of volume $V_{solid}$, $r$ can be expressed as 
\begin{equation}
r = \left(V_{solid} \frac{3(1-\widehat{\rho})}{4 \pi \widehat{\rho}}\right)^{1/3},
\end{equation}
so that the pore potential, eq. (\ref{eq:SurfacePotential}),  becomes
\begin{equation}
\label{eq:PorePontialFull}
   \psi_{pore} =\frac{\gamma_s}{\rho_{fd}V_{solid}^{1/3}} 4 \pi \left(\frac{3(1-\widehat{\rho})}{4\pi \widehat{\rho}} \right)^{2/3}. 
\end{equation}
Finally, a derivative of eq. (\ref{eq:PorePontialFull}) with respect to $\widehat{\rho}$ yields the sintering stress as
 \begin{equation}
 \label{eq:SinteringStressDerived}
 \sigma_s = 
\underbrace{\frac{8 \pi}{3}  \left( \frac{3}{4\pi} \right)^\frac{2}{3}}_{\approx 3.224} 
\frac{\gamma_s}{V_{solid}^{1/3}} 
\left(\frac{\widehat{\rho}}{1-\widehat{\rho}} \right)^{1/3},
\end{equation}
where ${V_{solid}^{1/3}}$ can be regarded as a length scale, equal to the particle diameter.
\section{Visco-Plasticity}
\label{sec:Plasticity}

The existence of a yield function is assumed, depending on the effective stress $\hat{\bm{\sigma}}$, which is the thermodynamic dual of the plastic strain rate, and on the internal variable $\widehat{\rho}$  
\begin{equation}
 \label{eq:YieldFunction}
 \mathscr{F}=\mathscr{F}( \hat{\bm{\sigma}}, \widehat{\rho}, T).
\end{equation}
The associative flow rule is assumed \cite{Doghri2000,Lubliner2005} involving the non-negative plastic multiplier $\dot{\lambda}$ 
\begin{equation}
 \label{eq:FloRulePerfectPlasticity}
 \dot{\bm{\epsilon}^p} = \dot{\lambda} {\bm{Q}}, 
\end{equation}
where the unit yield function gradient is 
\begin{equation}
{\bm{Q}} = \frac{\frac{\partial \mathscr{F}(\hat{\bm{\sigma}},\widehat{\rho}, T) }{\partial \hat{\bm{\sigma}} }}{\left\| \frac{\partial \mathscr{F}(\hat{\bm{\sigma}},\widehat{\rho}, T) }{\partial \hat{\bm{\sigma}} } \right\|}.
\end{equation}
For rate-dependent problems the coefficient $\dot{\lambda}$ can be replaced with an overstress function, as described in \cite{Perzyna1963}. 
A possible choice for the overstress function is the yield function divided by the viscosity ($\eta_v$). The expression for the strain rate becomes:
\begin{equation}
\label{eq:ViscoplasticFlowRule}
\dot{\bm{\epsilon}^p} = \frac{1}{\eta_v} \left<\mathscr{F}\right> {\bm{Q}}.
\end{equation}

\subsection{Yield Function} \label{ch:Triaxial}

The Bigoni-Piccolroaz (\lq BP' in the following) yield function \cite{Bigoni2004,Piccolroaz_2009} is defined by seven parameters and displays the necessary flexibility 
to describe a wide range of material behaviours and, in particular, can excellently fit the material behavior during the different states of sintering.
Therefore, the sintering process can be simulated from the beginning, when the material is in a granular form, and then during the firing, up to the finished ceramic piece.
The yield function is described by
\begin{equation}
 \label{eq:BPYield}
 \mathscr{F}(\bm{\sigma}, M,m, p_c, c, \alpha_{bp}, \Theta_c,\gamma_{bp},\beta_{bp}) =F(p,M,m,p_c,c,\alpha_{bp}) +q \, g(\Theta_c,\gamma_{bp},\beta_{bp}),
\end{equation}
where 
\begin{equation}
 \label{eq:BPyieldAdditionaltr1}
 p  = -\tr \bm{\sigma}/3,~~~~
 q  = \sqrt{\frac{3}{2}\dev \bm{\sigma} \cdot \dev \bm{\sigma}} ,
 \end{equation}
($\dev$ denotes the deviator part) and
 \begin{equation}
 \label{eq:BPyieldAdditionaltr2}
 \begin{aligned}
 F(p,M,m,p_c,c, \alpha_{bp}) &= -M p_c \sqrt{[\Phi_{bp}-(\Phi_{bp})^m][2(1-\alpha_{bp})\Phi_{bp}+\alpha_{bp}]}, \\
 ~~~  \Phi_{bp}&= \frac{p+c}{p_c+c}, \\
 g(\Theta_c, \beta_{bp}, \gamma_{bp}) &= \cos \left[ \beta_{bp} \frac{\pi}{6} - \frac{1}{3} \cos^{-1} (\gamma_{bp} cos ( 3 \Theta_c)) \right],
 \end{aligned}
 \end{equation}
in which the Lode angle $\Theta_c$ is defined as
\begin{equation}
\Theta_c = \frac{1}{3} \cos^{-1} \left(\frac{9\, \tr (\dev \bm{\sigma})^3}{2\,q^3}  \right) .
\end{equation}

Parameters $\alpha_{bp}$ and $M$ can be used to adjust the yield function to materials with internal friction.
Parameters $\beta_{bp}$ and $\gamma_{bp}$ determine the shape of the yield surface in the deviatoric plane, so that for both simplicity and lack of experimental data they are set to be zero, 
which means that Lode angle dependence is excluded (note that for triaxial compression the Lode angle $\Theta_c$ is equal to $\pi/3$, while it is zero for triaxial extension).
 The cohesion $c$ is the hydrostatic yield strength in extension whereas $p_c$ is the yield strength in hydrostatic compression.

\section{Micromechanics for the evolution of yield}
\label{sec:MicroModel}

To model the compaction behavior of a green body during the cold forming process, the following micromechanic assumptions are introduced, which allows the 
determination of the hardening parameters $p_c$, $c$, and $M$ (dictating the shape of the yield function), as functions of the internal variable $\widehat{\rho}$. 
It should be noted that the determination of the evolution of parameter $p_c$ with the inelastic density $\widehat{\rho}$ is equivalent to the determination 
of the theoretical compaction curve for a granulate material. 
This material is assumed to consist of elastic perfectly plastic particles, idealized as cylindrical (initially with circular cross section) in a two-dimensional approximation and obeying the Tresca yield criterion, that are equal in size and ordered in a centered cubic geometry, a disposition 
maintained fixed during sintering. The results of the two-dimensional approach will be tested against numerical simulations for a cubic disposition of 
perfectly-plastic and initially spherical particles, obeying the von Mises criterion. 
The particles thus yield all simultaneously, so that to predict the collapse load for this ensemble of particles, the upper bound theorem of limit analysis is applied.

\subsection{Plane strain upper bound for the determination of the compaction curve}
\label{sec:UpperBound2D}
A plane strain problem is considered, so that the ceramic particles are assumed in the form of circular cylinders, a geometry which may seem unrealistic, but has been proven to yield quite reasonable results \cite{Barbara1992,Argani2016}.
Employing the upper bound theorem of limit analysis \cite{Chen2008}, a collapse mechanism has to be assumed, which,  
even if it does not correspond to reality, it provides a simple evaluation of the dissipation that occurs within the body. 
Because the problem is reduced to two-dimensions, the particles are treated as planar figures, initially circular and later deforming and developing contacts through lines.  

\begin{figure}[htp]
 \centering
 \includegraphics[width=.99\textwidth]{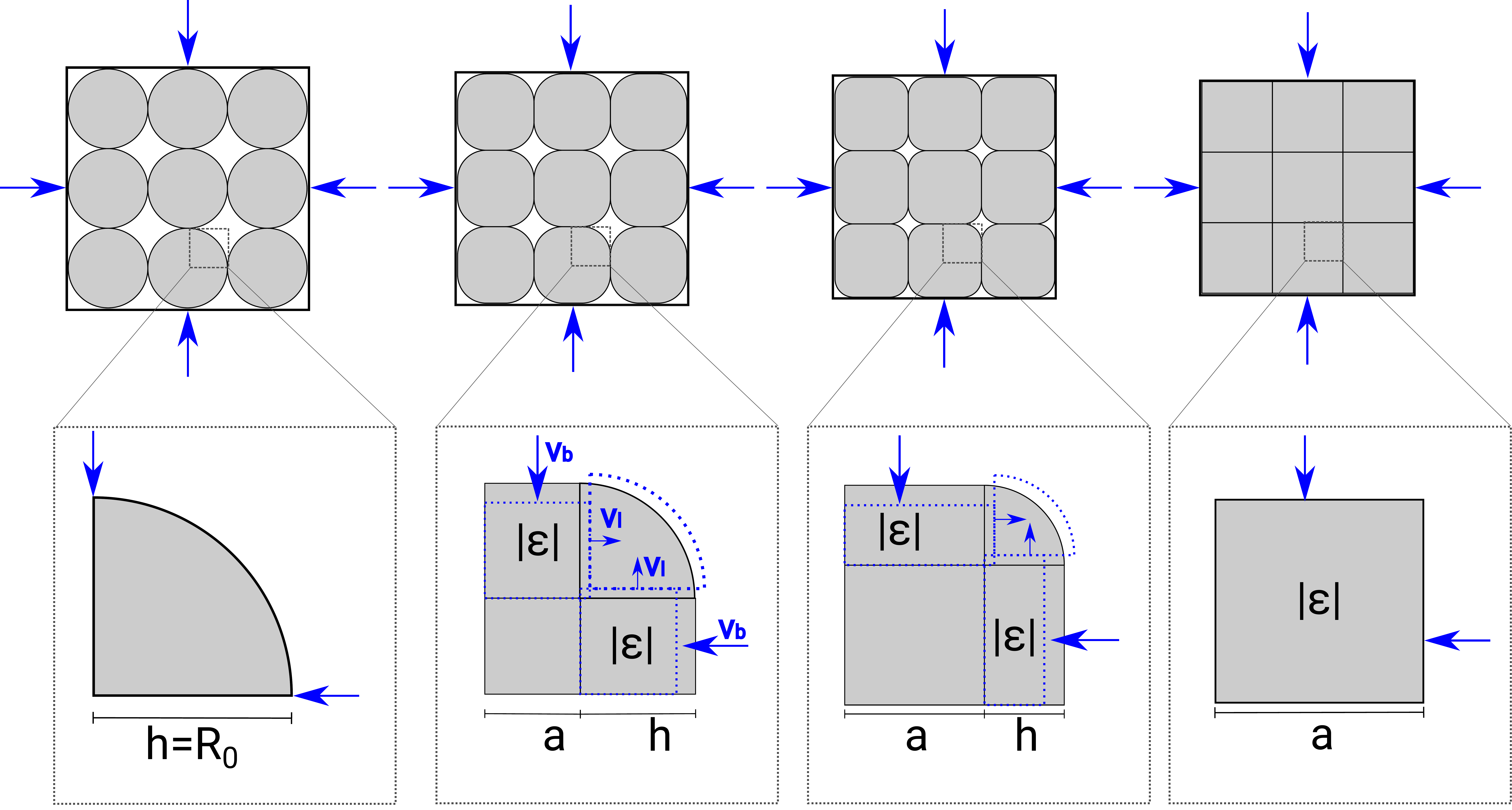}
 \caption{Assuming that the granules behave as rigid-perfectly-plastic materials, the isostatic compaction curve can be calculated using the upper bound technique of limit analysis. A representative volume element of an idealized 2D granular arrangement is shown in the figure at various stages of the compaction. 
 As the boundaries of the RVE are displaced with the velocities $v_b$, the RVE shrinks. 
 A section view of the assumed collapse mechanism, for a quarter of the circular particle, is reported in the lower line, where the dashed lines sketch the particle boundaries subject to a displacement rate.}
  \label{fig:LimitAnalysis2DHydv2}
\end{figure}

A collapse kinematics is assumed, in which rigid parts are subject to a pure rigid-body translation, while deformable parts are subject to a pure compressive strain, so that the latter (denoted by $|\epsilon|$ in Figure \ref{fig:LimitAnalysis2DHydv2}) are located on the surface of the particle in contact with the neighboring particles and the former are localized at the circular corners. The plastic mechanism has two axes of symmetry (vertical and horizontal) so that only a quarter of the particle 
is sketched and the two deformable parts suffer the same strain and dissipate the same power. 

The plastic dissipation is generated by the strain rate produced under uniaxial stress in two equally deforming blocks (denoted by $|\epsilon|$ in Figure \ref{fig:LimitAnalysis2DHydv2}) and their sliding against: (i.) a square rigid block located at the center of the particle and (ii.) other two rigid blocks having the shape of a quarter of circle. 

For an equibiaxial (vertical and horizontal) compression load $P$, which simulates mechanical compaction, the \lq external' rate of energy dissipation $\dot{W}$ can be written as 
\begin{equation}
 \label{eq:Limitanalysis1}
 \dot{W} = 2 P v_b ,   
\end{equation}
where $v_b$ is the velocity of the boundary in contact with the neighboring particles. 

The rate of internal power (denoted by $D$) is the sum of three dissipation sources, $D_{comp}$, $D_{slb}$, and $D_{tr}$, all related to the strain rate of the blocks denoted with the label $|\epsilon|$ in Figure \ref{fig:LimitAnalysis2DHydv2}. These blocks are subject to a linear velocity field preserving incompressibility, so that, assuming a $x_1$-$x_2$ reference system, with $x_2$ parallel to $v_b$
\begin{equation}
v_1 = \frac{v_l}{a} x_1,  ~~~ v_2 = \frac{v_b}{h} x_2, 
\end{equation}
where $v_l$ is the \lq lateral' velocity of the block, so that incompressibility requires $v_b a =-v_l h$.
Therefore, the three sources of dissipation rate can be calculated as follows. 

\begin{itemize}

\item Internal dissipation due to the strain rate of the rectangular blocks $|\epsilon|$
\begin{equation}
 \label{eq:DissipationCompressive}
 D_{comp} = 4 k v_b a, 
\end{equation}
where $k$ is the limit yield stress under shear (equal to 1/2 of the limit uniaxial stress). 

\item Dissipation due to the sliding of the blocks $|\epsilon|$ against the rigid square block at the center of the particle and against the quarter of circle rigid element
\begin{equation}
 \label{eq:DissipationSliding}
D_{slb} = 2 k \left( \int_0^a{v_1 \,} \, dx_1 + 
                     \int_0^h{v_2 \,} \, dx_2 \right) =  v_{l} a k + v_b h k.
\end{equation}

\item Dissipation due to rigid translation of the quarter of circle rigid element, occurring with sliding against one of the deforming blocks $|\epsilon|$ 
\begin{equation}
 \label{eq:DissipationTranslation}
D_{tr} = 2 v_{l} h k .
\end{equation}
\end{itemize}

In conclusion, summing up all the contributions from different dissipation sources, the internal expended power can be written as 
\begin{equation}
D = k v_b\left(6 a +a^2/h + h\right). 
\end{equation}

Eventually, the limit load $P$ can be expressed as a function of the yielding shear stress $k$ and of the current geometry (specified by $a$ and $h$) as
\begin{equation}
    P = k \left( 3a +\frac{1}{2} \frac{a^2}{h} + \frac{1}{2} h  \right) ,
\end{equation}
corresponding to the limit pressure that the particle characterized by the dimensions $a$ and $h$ can sustain. 

The width of the compressed part $a$ and its height $h$ can be expressed as functions of the relative density and the current 
side $R=a +h$ 
of the unit cell (initially equal to the radius $R_0$), so that, 
keeping into account incompressibility, the following expression is found
\begin{equation}
\begin{aligned}
 \label{eq:Density2dUc}
\widehat{\rho} &= \frac{A_0}{A_{UC}}=\frac{2ha+a^2+\frac{1}{4} \pi h^2}{R^2}  \\ &=\frac{2h(R-h)+(R-a)^2+\frac{1}{4} \pi h^2}{R^2},
\end{aligned}
\end{equation}
where $A_{UC}$  denotes the area of the unit cell containing the particle of area $A_0$. 
An expression for $R$ can also be found using the incompressibility constraint, and thus the fact that the initial area of the two-dimensional particle remains constant:
\begin{equation}
 \label{eq:RfromIncomp}
 \widehat{\rho} =\frac{\frac{1}{4} \pi R_0^2}{R^2}, ~~~
  \Rightarrow ~~~ R = \frac{R_0}{2} \sqrt{ \frac{\pi}{\widehat{\rho}}}.
 \end{equation}
Equation (\ref{eq:Density2dUc}) can be solved for $h$, so that using eq. \ref{eq:RfromIncomp} yields 
\begin{equation}
 \label{eq:hsol}
 h = \sqrt{\frac{\pi R_0^2 (1- \widehat{\rho})}{\widehat{\rho} (4 - \pi)}}
\end{equation}
and therefore the contact length $a$ results as
\begin{equation}
 \label{eq:asol}
a = R - h. 
\end{equation}

The hydrostatic yield stress $p_c$ is given by the force per unit length $P$ divided by $R$, so that the following representation of the compaction curve is obtained:

\begin{equation}
 \label{eq:PtimesSurface}
 p_c =  k \frac{\sqrt{\pi} \left(-8+ \left( 12 + \pi -16 \widehat{\rho} \right) \sqrt{\frac{\widehat{\rho}-1}{\pi-4}} +8 \widehat{\rho} \right)}{8\widehat{\rho}(\widehat{\rho}-1)} .
\end{equation} 
Eq. (\ref{eq:PtimesSurface}) describes the compaction curve of a granular material and is depicted in Figure \ref{fig:ComparisonOfPcModels}. 

\begin{figure}
  \centering
  \includegraphics[width=0.9\textwidth]{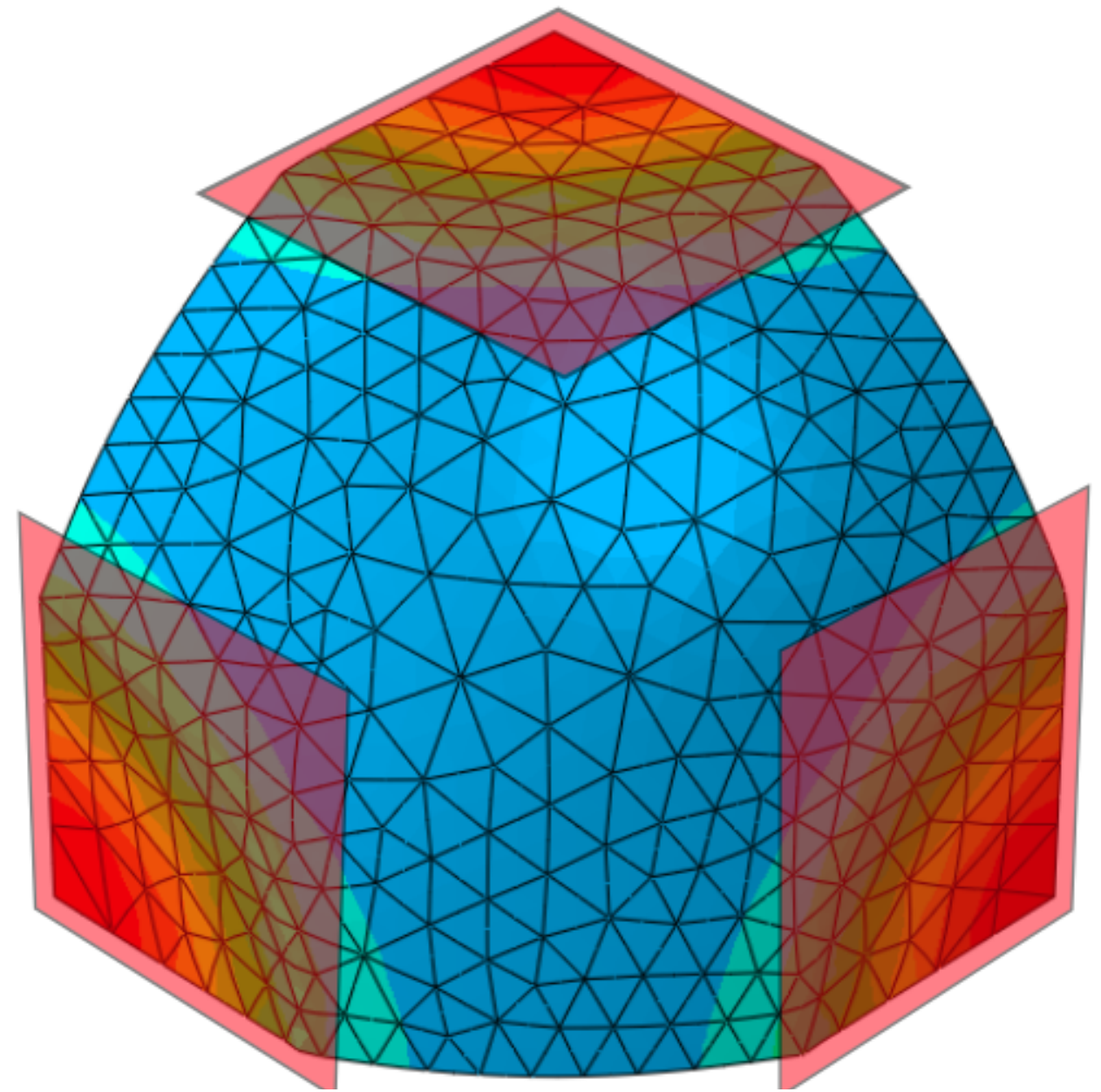}
  \caption{A finite element simulation of a cubic disposition of initially sherical particles, where, due to symmetry, only one eight of one particle is considered. 
  The particle is pressed into a cubic shape by three analytical contact surfaces (shown as translucent red surfaces) that move towards the center of the cell, with imposed displacements.
  Incompressible hybrid elements are used together with a neo-Hookean material model for the elastic part, while 
  the plastic behaviour is modelled with von-Mises perfect plasticity.}
  \label{fig:FEMParticle}
\end{figure}

\begin{figure}
  \centering
  \includegraphics[width=0.9\textwidth]{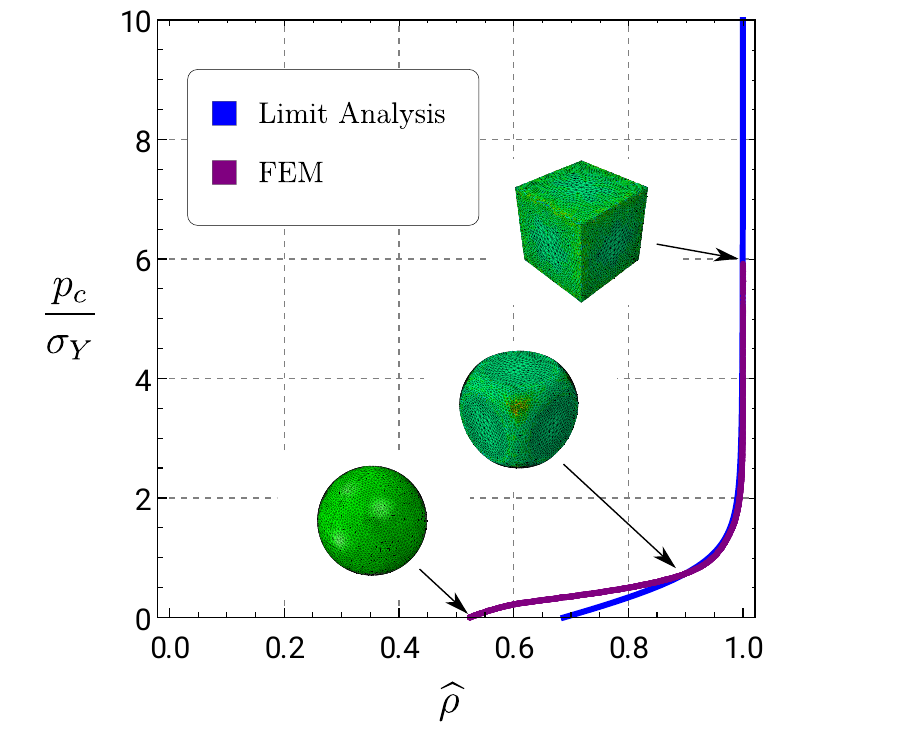}
  \caption{Comparison between the two compaction curves obtained from a two-dimensional application of the upper bound theorem of limit analysis and a three-dimensional finite element simulation of the unit cell shown in Figure \ref{fig:FEMParticle}, representative of the behavior of a cubic disposition of initially spherical particles. 
	A deformed particle at different levels of density is shown. 
  } \label{fig:ComparisonOfPcModels}
\end{figure}

Formally, Eq. (\ref{eq:PtimesSurface}) merely represents an upper bound, calculated for a plane strain situation, which may be believed  to be far from reality, so that an assessment of the capability of eq. (\ref{eq:PtimesSurface}) to correctly describe a compaction curve is provided through a comparison to a Finite Element simulation involving a three-dimensional distribution of initially spherical particles. 
The simulation was performed with perfectly-plastic particles, ordered in a simple cubic geometry. Initially, the contact between spheres occurs at a point, but due to the strain, the contact boundary becomes circular and towards the very end of the compaction tends to become square.
This periodic particle arrangement can be modelled using only one eighth of a sphere, due to symmetry. 
A FEM model was built in Abaqus 6.13, where displacement was imposed through analytical rigid body elements, sketched as colored planes in Figure \ref{fig:FEMParticle}, where a unit cell containing a deformed particle is shown. 
The material obeys von Mises plasticity, together with a neo-Hookean description for the elastic part, with a ratio of Young modulus over yield stress of $100000$, 
to approximate the rigid-plastic limit.
It can be seen that the solution of the limit analysis is, at low densities (about 0.8 and less), remarkably different 
from the numerical solution of the 3D problem, which is 
a direct consequence of the fact that the initial density for the 2D problem is different than that pertinent to the 3D arrangement. In fact, the relative density in plane strain is about 78\%, while this density decreases to about 52\%, in a cubic disposition of spherical particles. Moreover, a typical ceramic powder (as that later used for experiments, section \ref{sec:experiments}) has an initial density of about 38\%. A very simple way for correcting the discrepancy between the plane strain upper bound estimate and the values typical of ceramic powders is to introduce a correction factor $\zeta$ multiplying $h$, so that the initial value of $p_c$ becomes correct even for initial densities of 38\% and therefore the following \lq modified' value for $h$ is proposed 
\begin{equation}
 \label{eq:hsolmod}
 h_{mod} = \zeta \, R_0 \,\sqrt{\frac{\pi (1- \widehat{\rho})}{\widehat{\rho} (4 - \pi)}}, 
\end{equation}
where $\zeta=2.7$ provides the best fit to the material considered in the experiments. 
A substitution of eq. (\ref{eq:hsolmod}) into eq. (\ref{eq:Limitanalysis1}) yields an analytical expression for the compaction behaviour 
\begin{equation}
\label{eq:MicromechPc2}
 p_c (\widehat{\rho},\sigma_m) = \sigma_m \, \frac{-530+(619+25\pi-719\widehat{\rho}) \sqrt{\frac{\widehat{\rho}-1}{\pi-4}}+530\widehat{\rho}}{210\sqrt{3}(\widehat{\rho}-1)} .
\end{equation}
where $\sigma_m= k\sqrt{3}$ is the uniaxial stress for yielding. 
Note that eq. (\ref{eq:MicromechPc2}) provides an analytical description for the compaction curve of a granulate, which will be referred as the \lq modified limit analysis solution'.

\begin{figure}
  \centering
  \includegraphics[width=0.9\textwidth]{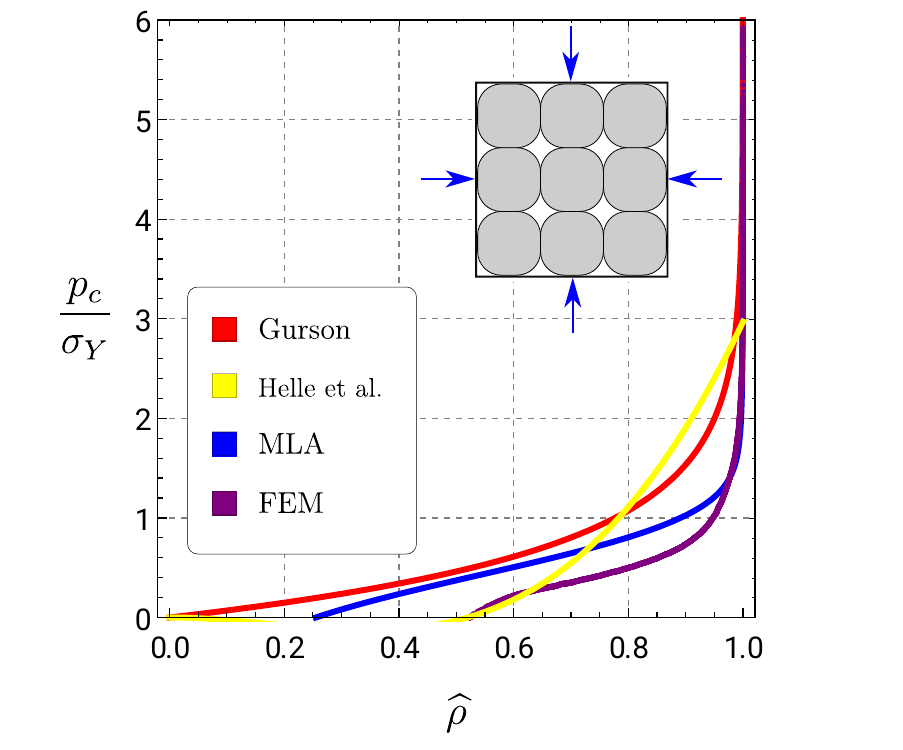}
  \caption{Comparison of the Gurson (\cite{Gurson1977}) and Helle/Fleck \cite{Barbara1992,Helle1985} models (for the compaction curve of a porous material) with 
	the modified upper bound solution, \lq MLA', eq. (\ref{eq:MicromechPc2}), and the three-dimensional finite element simulation, \lq FEM', of the unit cell shown in Figure \ref{fig:FEMParticle}.
  } \label{fig:ComparisonOfPcModelsEd}
\end{figure}
Figure \ref{fig:ComparisonOfPcModelsEd} shows the Gurson \cite{Gurson1977} and Helle/Fleck \cite{Helle1985,Barbara1992} models compared with the previously-described FE simulation of a three-dimensional cubic disposition of initially spherical particles (\lq FEM' in the figure) and the \lq modified limit analysis solution' (\lq MLA' in the figure). 
The modified limit analysis solution lies below the curves corresponding to the Gurson and Helle/Fleck models, at high porosity. 
It needs to be mentioned that the Gurson model was originally developed for materials with high density, where the modified upper bound solution and the numerical simulation almost coincide, whereas the model presented by Helle \cite{Helle1985} (and used also by Fleck \cite{Barbara1992}) was intended for cases of 
high porosities, where predictions of this model are consistent with the modified upper bound and the numerical simulation.
\subsection{Cohesive strength under tension}
The values for $p_c$ (under compressive load) have been determined as a function of $\widehat{\rho}$. 
The cohesion strength $c$ can also be evaluated as a function of the 
internal variable $\widehat{\rho}$ from the following simple micromechanical model. 
Simulations performed with the unit cell sketched in a deformed state in Fig. \ref{fig:FEMParticle} show that the contact area between initially spherical grains in a cubic geometry is a complex function of the applied pressure, so that an initially circular contact area evolves towards a square contact shape in the limit of a fully dense material.
Therefore, the estimation of the contact areas between grains in a real powder subject to cold pressing is awkward. Nevertheless, following \cite{Helle1985}, an approximation for the normalized (and dimensionless) contact area $A_c$ between grains can be assumed as
 \begin{equation}
 \label{eq:UnitContactArea}
 A_c= \frac{4 \pi}{12} \frac{\widehat{\rho}-\widehat{\rho}_0}{1-\widehat{\rho}_0}, 
 \end{equation}
so that the stress needed to produce yielding under hydrostatic tension (which is the definition of the cohesion $c$) is given by the elementary formula
 \begin{equation}
  \label{eq:CoehsionInMicroModel}
  c = \sigma_m \, A_c,
 \end{equation}
where $\sigma_m$ is the yielding stress in uniaxial tension of the grains. Note that the normalized contact area goes to unity as $\widehat{\rho}$ approaches one.

\subsection{Strength under shear for the determination of parameter M}

Using again relation (\ref{eq:UnitContactArea}), the parameter $M$ of the BP yield surface can be expressed as a function of the internal variable $\widehat{\rho}$. In particular, it is assumed that at failure under pure shear ($p=0$ so that $\bm{\sigma}= \dev \bm{\sigma}$) the deviatoric component of the stress tensor becomes,  according to the von Mises criterion,  
$\sigma_m A_c/3$, so that the stress invariant $q$ at failure (denoted by $q_s$) reduces to $\sigma_m A_c$. Therefore, the following relation is obtained (assuming $\beta=\gamma=0$) for the parameter $M$ 
   \begin{equation}
  \label{eq:MFromMicroModel}
M= \frac{\sqrt{3}\sigma_m A_c}{p_c \,2\sqrt{[\frac{c}{p_c+c}-(\frac{c}{p_c+c})^m][2(1-\alpha_{bp})(\frac{c}{p_c+c})+\alpha_{bp}]}}.
 \end{equation}

\section{Influence of Temperature}

The consolidation process of a ceramic body is activated at high temperatures, where 
the material exhibits a strong thermal softening. 
Also, the viscosity changes drastically with temperature, so that viscoplastic processes occur at much higher velocities, when the temperature is high.

\subsection{Temperature evolution}
\label{sec:TempEvol}
In the framework proposed by Simo and Miehe \cite{Simo1992}, the temperature evolution is governed by the Fourier law of heat conduction, together with a maximum dissipation postulate.
In the absence of internal heat sources, the first law of thermodynamics is 
 \begin{equation}
  \label{eq:EvolutionequationsHeat}
  \dot{T}=\frac{1}{\rho c_h}\Big(\textrm{div }(k\nabla T) +\mathcal{H}+\mathcal{D}_{mech} \Big), 
 \end{equation}
where $c_h$ is the specific heat at constant values of strain and internal variables, eq. (\ref{ci}), 
$k$ the thermal conductivity, eq. (\ref{fou}),
$\mathcal{D}_{mech}$ is the mechanical dissipation power, 
\begin{equation}
\label{eq:plasticheating}
\mathcal{D}_{mech} = \bm{\sigma}\cdot\dot{\bm{\epsilon}}_p+R (\widehat{\rho})^\cdot=\hat{\bm{\sigma}}\cdot\dot{\bm{\epsilon}}_p, 
\end{equation}
and $\mathcal{H}$ is the non-dissipative elastic-plastic heating
 \begin{equation}
  \label{eq:NonDissipativeElasticPlasticHeating}
  \mathcal{H} =  T \left[ \frac{\partial^2 \psi}{\partial T \partial \bm{\epsilon}_e} \cdot \dot{\bm{\epsilon}}_e+\frac{\partial^2 \psi}{\partial T \partial \widehat{\rho} } (\widehat{\rho})^\cdot \right].
 \end{equation}
 The first term inside the parenthesis describes the piezocaloric effect. 
 The second part is a coupling term between the temperature and the internal variable (relative density).
 It is common for plasticity models \cite{Simo1992} to replace the mechanical dissipation term with
 \begin{equation}
  \label{eq:TaylorQuinney}
 \mathcal{D}_{mech}=\mathcal{X} \hat{\bm{\sigma}} \cdot \dot{\bm{\epsilon}},
 \end{equation}
where $\mathcal{X}$ is sometimes called Quinney-Taylor coefficient and lies between $0.85$ and $0.95$. 
This factor was introduced as it was found in experiments \cite{Taylor1934} that the heating associated with the plastic flow is roughly 0.9 times the rate of plastic working. 
If the plastic strain becomes very large, the Quinney-Taylor factor approaching unity, so that eq. (\ref{eq:TaylorQuinney}) reduces to eq. (\ref{eq:plasticheating}). 
However, the mechanical dissipation plays only a negligible role in high temperature applications. 
In the case of ceramic sintering, a substantial amount of heat is transferred to the system from outside (for instance, the ceramics is left 30 min in the oven at about 1200$^\circ$ C), whereas the plastic deformation of the body occurs at a relatively slow pace.
This heat transfer is far larger than the heating introduced by the piezocaloric effect or the plastic heating. 
It is therefore reasonable to neglect these terms (the mechanical dissipation and the non-dissipative elastoplastic heating) and assume $\mathcal{X}=0$, which leads to a system where temperature evolution is uncoupled from plastic flow. 
Then, eq. (\ref{eq:EvolutionequationsHeat}) reduces to:
 \begin{equation}
  \label{eq:EvolutionequationsHeatUncoupled}
  \dot{T}=\frac{1}{\rho c_h} \, \textrm{div }(k\nabla T).
 \end{equation}

\subsection{Temperature effect on the yield surface}

At high temperature, a significant thermal softening effect occurs, modelled through a variation of the parameter $p_c$, which is assumed to follow the rule
\begin{equation}\label{eq:ThermalSoftening1}
p_c (\widehat{\rho},\sigma_m, T)= f_T(T) \, p_c (\widehat{\rho},\sigma_m),
\end{equation}
with $f_T(T)$ being a steadily decreasing function of temperature, equal to $1+C_T$ at a reference temperature of $0^\circ C$ . 
For high temperatures, over about $800^{\circ}C$ and slow deformation rates, the material exhibits creep as dominant deformation mechanism \cite{Ashby1981,Hammond2004}, and its strength is assumed to become negligible, so that 
\begin{equation}
\label{ch:five:sec:5:eq4:1}
f_T(T) =  \left<1-\frac{T}{T_{C1}} \right>^{b_1}  + C_T, 
   \end{equation}
where the Macaulay bracket has been used and $T_{C1}=800^\circ C$ and $C_T=0.0001$, with $b_1=0.9$.

\subsection{Grain growth}

\label{sec:GrainGrowth}
The sintering process is usually accompanied by grain coarsening \cite{Rahaman2007}, which has some effect on the process kinetics and therefore has to be considered
in the modelling.
The grain growth occurring during sintering is assumed to follow an exponential law that models the evolution of the average grain size, as described in \cite{Wonisch2009} and \cite{Hillert1965}, namely
\begin{equation}
 \label{eq:ParticleSizeEvolutionLaw}
 \dot{R}= \frac{\gamma_b M_{gc}}{4R}, ~~~~
 M_{gc} = M_{gc0} \text{ exp} \left( {\frac{-Q_{gc}}{R_g T}} \right),
  \end{equation}
where $R_g$ is the universal gas constant (8.314 J/(mol K)), the constant  $M_{gc0}$ is taken from \cite{Wonisch2009} to be equal to 2.25 m$^2$ s/kg and $Q_{gc}$ is the activation energy for grain coarsening.
The value for the grain boundary energy $\gamma_b$ is also taken from \cite{Wonisch2009} and is equal to 1.10 J/m$^2$ .
The values of the activation energies, to be used for the model, can be obtained by fitting them to experimental results, with a procedure described in section \ref{sec:Calibration}.

\subsection{Viscosity}
\label{sec:viscosity}

The shrinkage of the ceramics under thermal load during sintering is essentially a time-dependent process, so that the previously developed time-independent  model has to be enhanced to describe rate-dependent effects, thus introducing viscosity.
There are several possibilities to introduce viscous behaviour, for instance using the Coble \cite{Coble1963} or the Nabarro-Herring \cite{Rahaman1986} creep model.
Olevsky \cite{Olevsky1998} introduces a model based on a dissipation potential that can be derived from a strain energy function. 
The viscous models proposed in \cite{Bordia1988, Rahaman1986} are compared with experimental results in \cite{Zuo2003, Zuo2004} and show relatively good fit. 
Thus, a viscous model is also used in the present study.
The dependence of the viscosity $\eta_v$ on the temperature, density and grain size is assumed to be multiplicative as
\begin{equation}
 \label{eq:DefineViscosity}
\eta_v =\eta_{v1} \cdot \eta_{v2}(R) \cdot \eta_{v3}(T),
\end{equation}
where $\eta_{v1}$ is a constant and $\eta_{v2}$ and $\eta_{v3}$ are respectively functions 
of the radius $R$ of the particles and of the temperature $T$.  
In particular, introducing the initial radius $R_0$ of the particles, the function $\eta_{v2}$ is assumed in the form
\begin{equation}
\label{eq:ViscDepOnR}
 \eta_{v2}(R)=\left(\frac{R}{R_0}\right)^w,
 \end{equation}
where the exponent $w$ is equal to 3 for grain boundary diffusion as the main type of inter-particle diffusion mechanism \cite{Rahaman1986}.
Finally, the temperature dependence in the law (\ref{eq:DefineViscosity}) is assumed to follow an Arrhenius type law (compare with \cite{Zuo2004}):
\begin{equation}\label{eq:ViscDepOnT}
\eta_{v3}(T)=e^{\frac{Q_E}{R_g T}},
\end{equation}
where $Q_E$ is the activation energy for the viscosity and $R_g$ the universal gas constant (8.314 J/(mol K )). 

In conclusion, the complete function governing the viscosity is 
\begin{equation}
\label{eq:ViscosityFunction}
\eta_v= \eta_{v1} \left(\frac{R}{R_0}\right)^w e^{\frac{Q_E}{R_g T}},
\end{equation}
which is used for the viscoplastic flow rule, eq. (\ref{eq:ViscoplasticFlowRule}).
The values of the constants were fitted against a sintering curve obtained from one dilatometric test, as described in section \ref{sec:Calibration}.

\section{Computer implementation}

The material model developed in the previous Sections was implemented in Abaqus 6.13 (Dassault Syst\`emes) through a UMAT routine created with AceGen \cite{Korelc2009a}. 
An implicit scheme was used for the yield function, as explained in \cite{Stupkiewicz2014}, to overcome difficulties linked to non-convexity, see \cite{brannon2010multi,Penasa_2014}.
The viscoplastic deformation is evaluated using a return-mapping algorithm (as described in \cite{Simo1992, Simo1998a, deSouza2008}), which was implemented taking advantage of the Automatic Differentiation technique \cite{Korelc2016, Stupkiewicz2014}, to obtain the derivatives (first and second) needed in the algorithm. 
A consistent tangent matrix for the local Newton method, usually employed in plasticity, was obtained (as described in \cite{Miehe2002,deSouza2008}), without explicitly calculating  cumbersome derivatives.
The viscosity is implemented using a Perzyna \cite{Perzyna1963} type approach, as described in Section \ref{sec:Plasticity}.
As the model uses logarithmic strains, and finite deformation, the `Nlgeom' option of Abaqus was used, which provides logarithmic strains as the input for the user material routine and uses finite deformation kinematics.
The calibration of the yield function, using nominal stress, is described in section \ref{sec:CalibrationAndValidation}. 
For the solution of the global scheme, a separated Newton algorithm was used, in which the coupling terms of the global jacobian matrix (the entries that are a result of differentiation with respect to both displacement and  temperature) are set to be zero.
This can be justified because the problem is weakly coupled (see Sec.~\ref{sec:TempEvol}), and, in addition, the computational cost results strongly reduced. 
The Newton technique was modified using a line search method, to improve convergence. 
Note that the general accuracy of the solution remains unchanged by using the separated solution scheme, while 
only the rate of convergence is slightly affected. 

The experimental setup, described in Section \ref{sec:experiments}, was modelled in Abaqus, where the mold and the stamp were modelled as analytical rigid bodies and the displacement was imposed on the stamp. To model the furnace environment, the temperature curve was imposed on the boundaries of the green body, obtained in an  initial step through the modelling of the cold powder compaction.

\section{Validation and calibration}
\label{sec:CalibrationAndValidation}

Experiments have been performed to calibrate and subsequently validate the material model. In particular, 
the calibration of the viscosity parameters of the model has been conducted with a dilatometric test, while the yield function parameters have been 
evaluated on the basis of available experiments \cite{Bosi2014}. 
Finally, validation of the model performance was obtained through a comparison between model predictions and the shrinking of a specially designed ceramic piece, measured during sintering.

\subsection{Calibration of the viscosity parameters}
\label{sec:Calibration}

\begin{figure}[htb]
 \centering
  \includegraphics[width=1\textwidth]{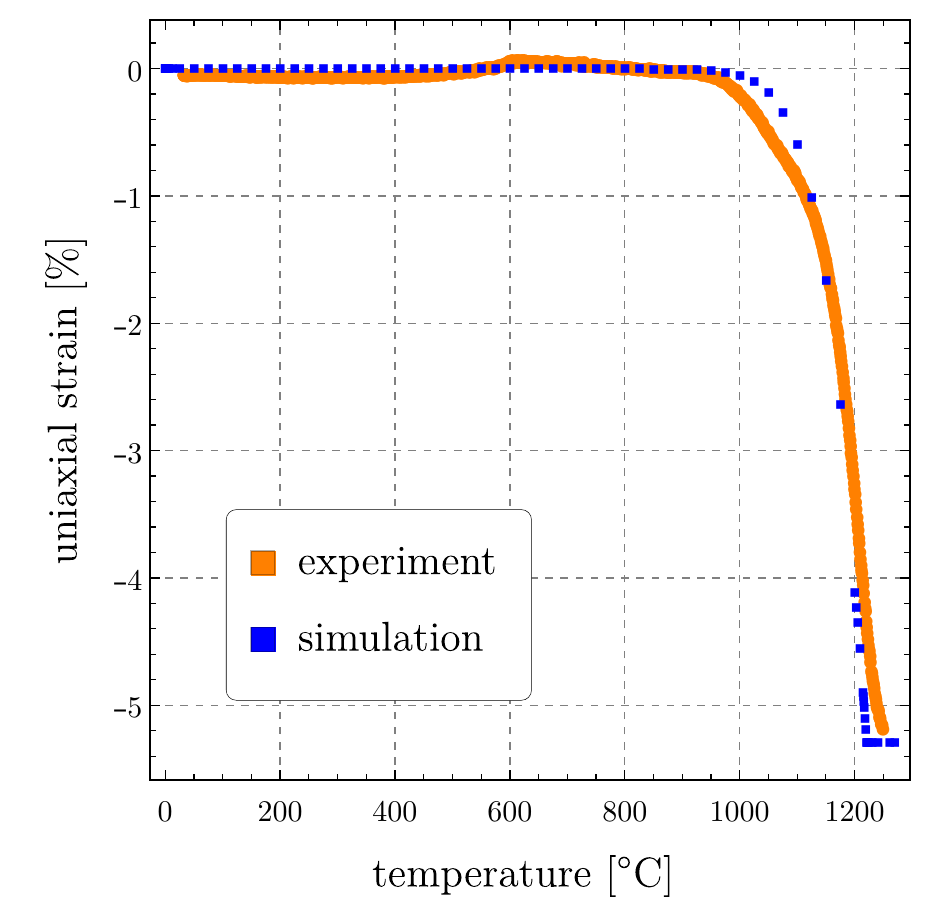}
  \caption{
	Activation energies and the viscosity constants have been determined with repeated numerical simulations of the experimental results obtained from a dilatometer test, until a good fit has been found (illustrated in the figure).}
\label{fig:DILTest}
\end{figure}

The sintering curve can be obtained using a dilatometer, a standard equipment in laboratories dealing with ceramics.
Using a special dilatometer and through  
a long series of  experiments, a method has been proposed  for experimentally determining the viscosities and activation energies for pure alumina \cite{Zuo2003,Zuo2004}. 
Instead of this complex procedure, the viscosity parameters of the proposed model were estimated through a comparison between results from dilatometer tests and 
numerical simulations in which only a single element was used. The temperature was assigned as a constraint on all nodes, following the temperature curve of the dilatometer. 
This procedure can be justified because of the very small dimensions of the specimen, which does not possess a large thermal capacity.
The values for the viscosity constant $\eta_{v1}$ and for the activation energies $Q_{E}$ and $Q_{gc}$ were found by running simulations with different values and iterating towards an optimal fit. 
This simplified approach was chosen because experimental values for the activation energies of traditional ceramics are not available.  
It was assumed that the activation energies for grain coarsening and viscosity are equal, an assumption based on the fact that these values have been found very similar for ceramic materials (but different from the powder addressed in the present study \cite{Zuo2004}). 
Experimental data for mechanical properties are rare, as their determination is not straightforward \cite{Gei2004} and often not of interest for the involved industries. 
However, a yield strength ranging between $125$ MPa and $250$ MPa is provided in \cite{Fragassa2014}, so that $\sigma_m=150$ MPa has been assumed.
Aluminum silicate spray dried powder (Sacmi I20087) was used, with a theoretical density (when pores are completely absent) of about 2.375~g/cm$^3$.
To obtain the sintering curve from the powder, a dilatometer (TAinstruments DIL-831) was used at a heating rate of 30$^\circ$C/min, equal to the heating rate employed in industrial furnaces. 
Data in Figure \ref{fig:DILTest} have been plotted after the strain due to thermal expansion was subtracted, to yield a graph that is corrected for thermal expansion, which 
was measured in a previous trial, with a completely sintered piece. 
The best fit was found for the viscosity constant $\eta_{v1}=1 \cdot 10^{-8} \text{ MPa} \cdot \text{s}$ and activation energies $Q_{E}=Q_{gc}$=340~kJ/mol. 

\subsubsection{Calibration of the yield function parameters}

\begin{figure}[htb]
 \centering
  \includegraphics[width=1\textwidth]{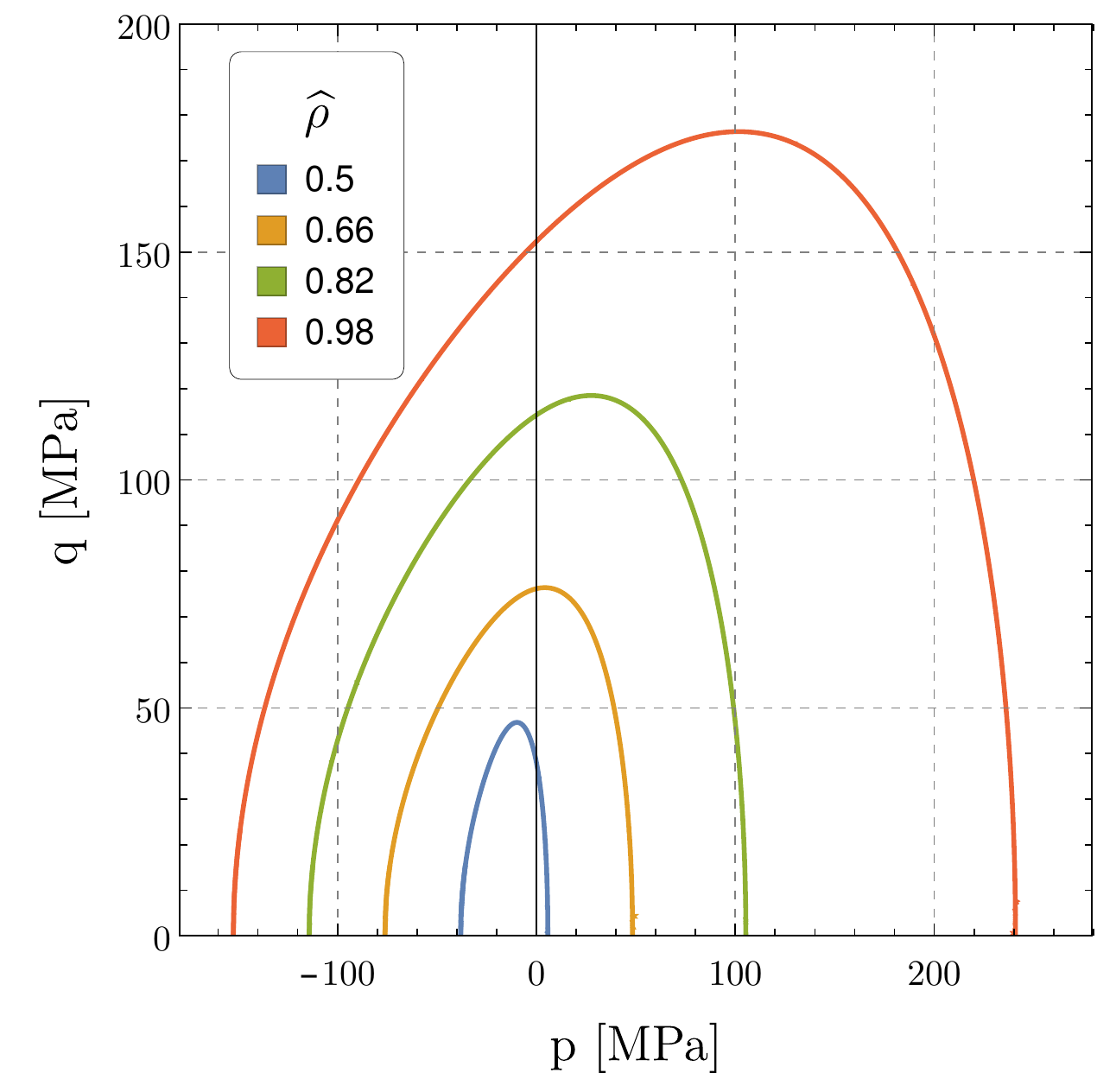}
  \caption{The BP yield surface evolution with respect to the relative density, for a porcellain stoneware ceramic (aluminum silicate spray dried powder). 
  The parameters defining the yield surface are $\alpha_{bp}=1$, $m=4.38$, $\gamma_{bp}=0$, $\beta_{bp}=0$, $\sigma_m=$150MPa}
\label{fig:BPVis}
\end{figure}

The BP yield function, eq. (\ref{eq:BPYield}), requires the determination of seven parameters, but   
using the previously developed micromechanical modelling, the number of unknown material constants can be reduced to four.
During powder compaction and sintering, in most of the cases the body is under a compressive load. Therefore, the lode angle-dependence is of minor importance and can therefore be 
neglected, so that parameters  $\beta_{bp}$ and $\gamma_{bp}$ are set to be zero.
The parameters $\alpha_{bp}$ and $M$ determine the shape of the yield surface in the $p-q$ plane. 
The value $m=4.38$ has been deduced from \cite{Bosi2014}, while $\alpha_{bp}$ is chosen to be equal to the unit, as in  the simple case of the modified Cam-Clay model \cite{Bigoni2004}. 
The size of the yield surface in the $p-q$ plane is determined by $p_c$, $c$ and $M$, which are functions of the relative density. 
The yield surface evolves as the density grows, so that this 
evolution is shown in Figure \ref{fig:BPVis}.

\clearpage

\subsection{Material parameters used for the examples}

In summary, the values of the parameters used in the simulations, defining the material behaviour and the contact conditions at the powder/mold interface, where a Coulomb friction law was assumed,  are listed in Table \ref{tabella}. 

\vspace{10 mm}

\begin{tabular}{l | r}
Material Parameters           & Values \\
\hline \hline
$R_0$: Initial particle radius            & 11.24 $\cdot 10^{-6}$m  \\ \hline
$\gamma_s$: Surface energy & 1.10 J/m$^2$ \\ \hline
Q$_{E}$: Viscous activation energy  & 354 kJ/mol\\ \hline
 M$_{gc0}$: Grain boundary mobility coefficient & 2.25 m$^2$s/kg\\ \hline
 Q$_{gc}$: Grain boundary mobility activation  & \\
Energy &354 kJ/mol   \\ \hline
$E$: Young's modulus  & 5000 MPa \\ \hline
$\mathcal{X}$: Taylor Quinney Coefficient  & 0 \\ \hline
$\sigma_m$: Compressive strength \\ of the fully dense material  & 150MPa  \\ \hline
$w$: Exponent in the viscosity law  & 3 \\ \hline
$R_g$: Gas constant  & 8.314 J/(mol K) \\ \hline
$T_{C1}$: Temperature constant \\ for thermal softening law & 800\degree C \\ \hline
$T_{C2}$: Constant for thermal softening law & 0.0001 \\ \hline
$b_1$: Constant for thermal softening law & 0.9 \\ \hline
$\rho_{0}$: inital density & 0.38 \\ \hline
$\eta_{v1}$: viscosity constant & $10^{-8}$ MPa$\cdot$s \\ \hline
$m$: BP Parameter & 4.38 \\ \hline
$\beta$: BP Parameter & 0 \\ \hline
$\gamma$: BP Parameter & 0 \\ \hline
$\alpha$: BP Parameter & 1 \\ \hline
$\mu$: Die/stamp  Coulomb friction coefficient & 0.4 \\ \hline
\label{tabella}
\end{tabular}

\newpage

\subsection{Experimental validation}
\label{sec:experiments}

\begin{figure}[ht]
\includegraphics[width=\textwidth]{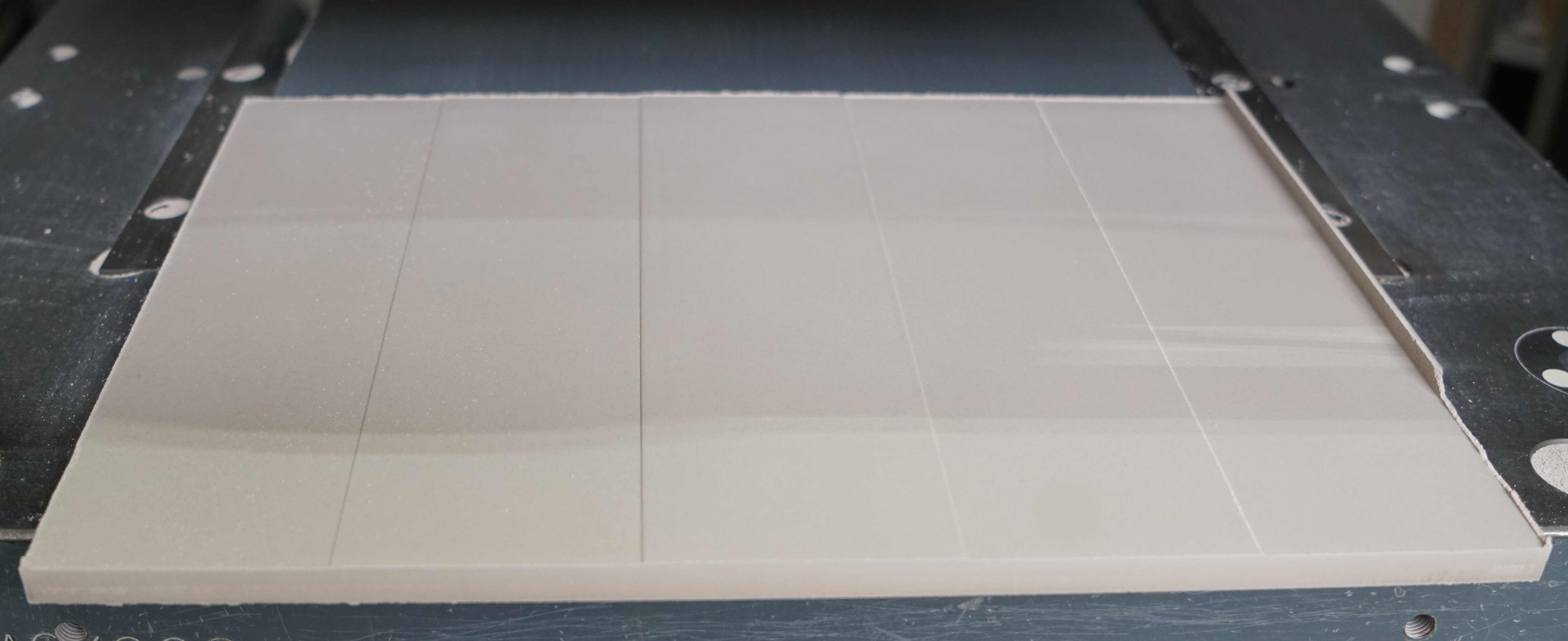}
 \includegraphics[width=0.8\textwidth]{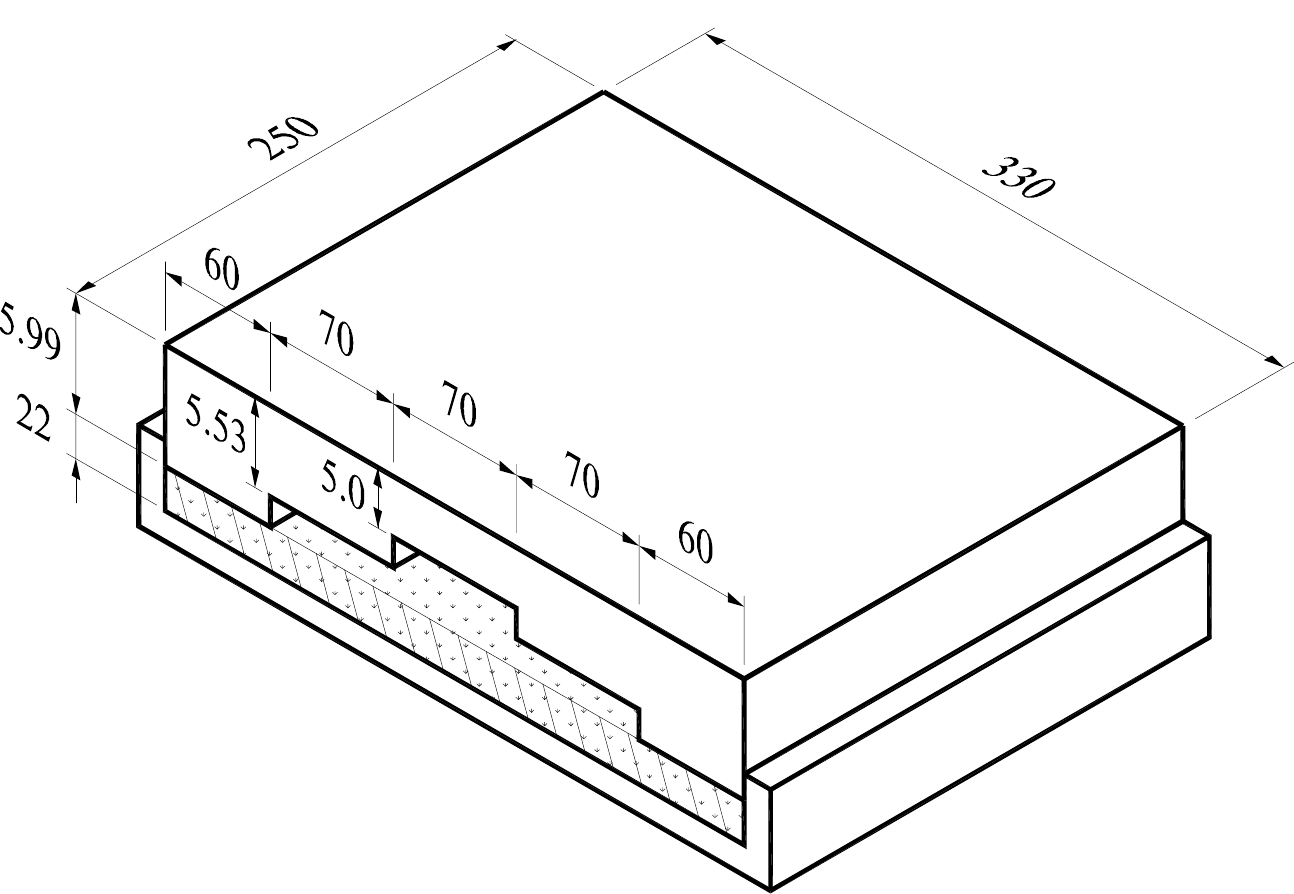}
\caption{
Upper part: Photo of the green body, with zones of different height and therefore density, used for sintering and density measurements. 
The green was formed with a tool, which was manufactured with three different heights, sketched in the lower part of the figure (not true to scale).}
 \label{fig:SketchPressing}
 \end{figure}

Validation of the model has been obtained by referring to the forming and subsequent sintering of a special green piece, namely, a profiled tile.
It is possible to introduce some complexity to a floor tile by using a special tool with varying heights (as sketched in Figure \ref{fig:SketchPressing}), 
thus obtaining a green with zones of different height and therefore density.
Before sintering, this profiled geometry has to be pressed from powder, using a tool that was manufactured for this specific purpose (Fig. \ref{fig:SketchPressing}, lower part). 
The dimensions before pressing were 330mm $\times$ 125mm, with an initial uniform thickness of 22mm.
The profiled tile was pressed with a symmetric tool, to avoid lateral loads during stamping, as sketched in Figure \ref{fig:SketchPressing}.

\begin{figure}[ht]
 \includegraphics[width=0.9\textwidth]{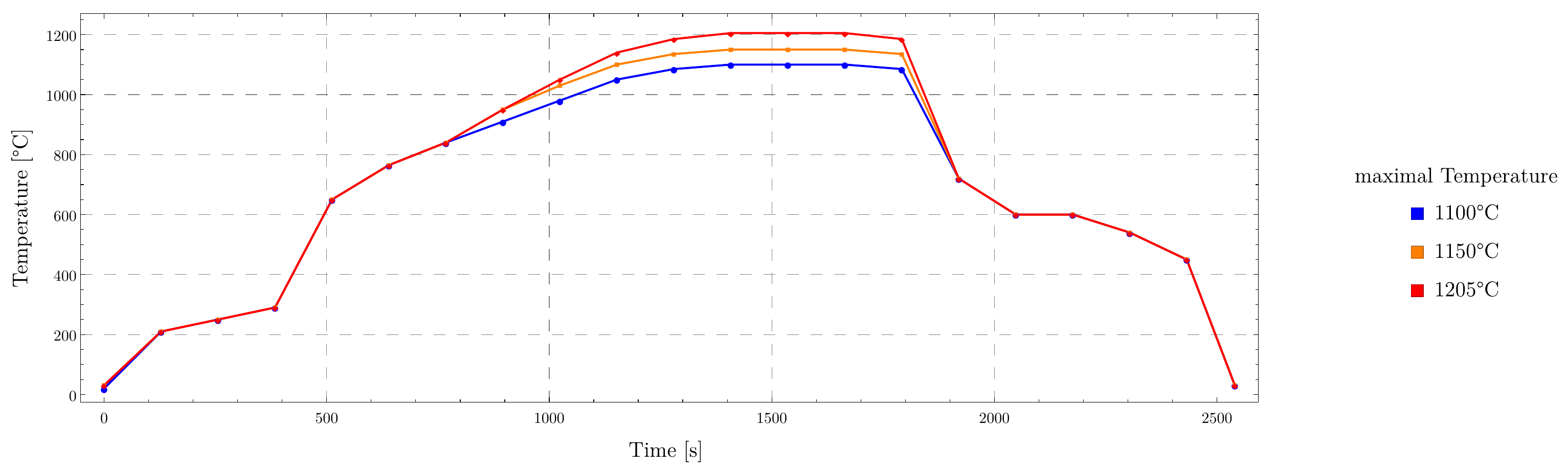}
\caption{Sintering was obtained by moving the green through a continuous oven (a \lq Sacmi Forni S.pA. EUP 130 was used) across different temperature zones, so that the time-temperature curve shown above is applied.}
 \label{fig:OvenTempCurves}
 \end{figure}

Twenty profiled tiles were formed and their density measured in the Laboratory of Sacmi (located at Imola, Italy). 
All the ceramic greens were found to be so closely similar in their density distributions, that it was decided to fire only 6 greens. 
Three groups of two greens were fired respectively at temperatures of 1100$^\circ$C, 1150$^\circ$C and 1200$^\circ$C, by adopting the heating cycle shown in 
Figure \ref{fig:OvenTempCurves}, using an oven available at the Sacmi Labs.

\begin{figure}[ht]
 \includegraphics[width=0.9\textwidth]{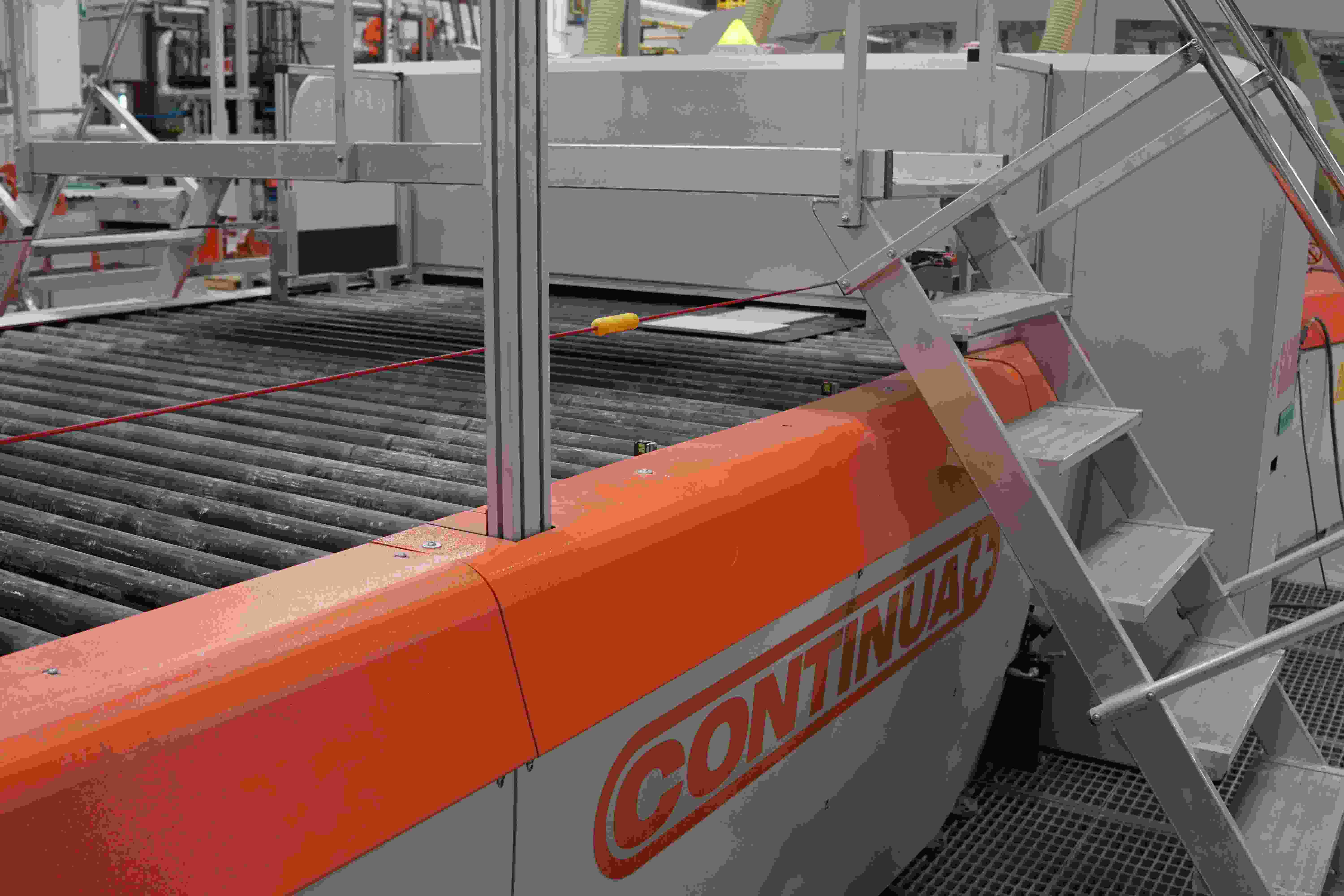}
\caption{Density measurements were performed with an X-Ray scanner from the \lq Sacmi CONTINUA+' line, shown in the photo. Note that a profiled tile is entering the scanner.}
 \label{fig:CTScanner}
 \end{figure}

The density variation before and after sintering was measured using the purpose-made X-ray Line Scanner (of the \lq Sacmi Continua+' line, manufactured by Microtec Srl, Bressanone, Italy, Fig. \ref{fig:CTScanner}) at the Sacmi  laboratory. While the thickness changes abruptly, creating a tile with three rather distinct heights (9.8mm, 10.2mm and 10.7mm), the density varies less abruptly, as can be seen in Figure \ref{fig:TileDensity20}. The density distribution was measured on pairs of nominally identical tiles, both subject to the same treatment. The densities in the pairs (before and after firing) was found so similar (the experimental data were almost superimposed), so that only one experiment for each pair is reported below.

 
\clearpage
\subsection{Simulation of the forming and sintering of the ceramic plate with different densities}
\label{sec:Implementation}

\begin{figure}[ht]
 \includegraphics[width=0.9\textwidth]{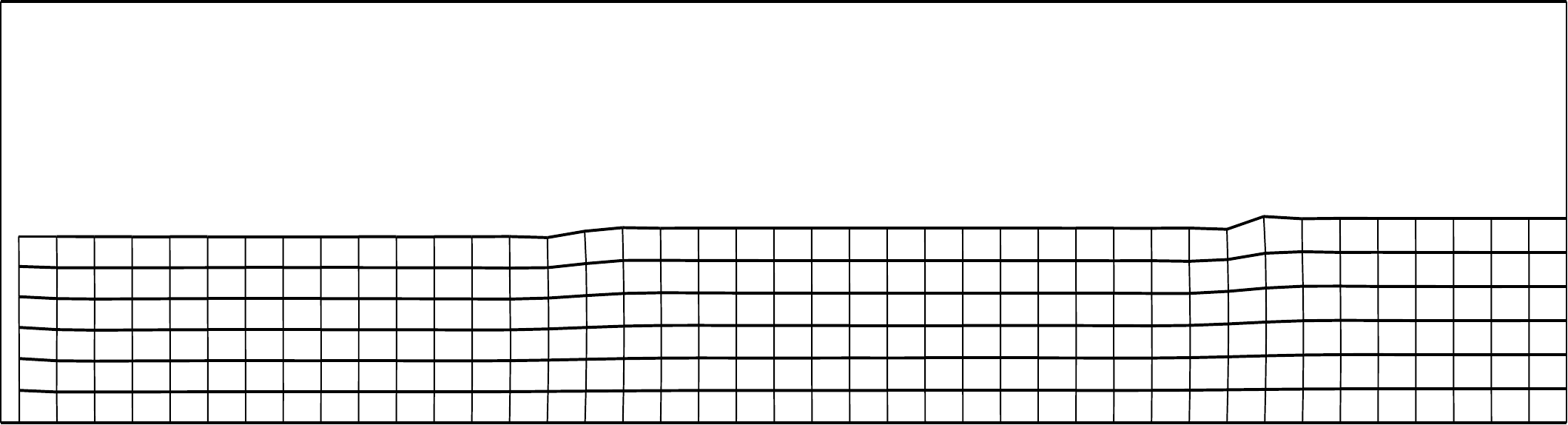}
\caption{The mesh of the modelled powder in its initial state (a rectangle, not meshed) and after pressing and subsequent release from the mold (with mesh). 
Just one half the piece was simulated, due to symmetry. 
The figure is scaled by 200\% in height direction, to make the contour change more visible.}
\label{fig:MeshesFromSide}
\end{figure}

The stamp and mould for powder forming were modelled as rigid bodies in contact with Coulomb friction (coefficient equal to 0.4). 
The powder was initally considered of uniform height (22mm) and homogeneous relative density (0.38), as in the experimental setup. 
The stamp is then displaced by $\Delta H$ = 12.6~mm, pressing the body in the desired shape. 
Powder pressing is usally modelled as a rate-independent process \cite{Piccolroaz_2006}, so that for the powder pressing part of the simulation, the viscosity was set to a constant and low value, to come close to the limit of rate-independent plasticity.
For the simulation of sintering, the viscosity model reported in section \ref{sec:viscosity} was used and the temperature curve of the oven was prescribed on the boundaries of the ceramic body.
The simulated geometry is shown in Figure \ref{fig:MeshesFromSide} (not to scale), where the undeformed piece (a rectangle, unmeshed) and the deformed mesh after pressing and subsequent mold release are reported. 
The figure clearly shows the large strain suffered by the powder during compaction and the presence of a modest spring back effect, which 
is limited in the lateral direction, because of the high aspect ratio of the sample. In particular, the tile increases in height after release, but decreases a little bit in width. 

\begin{figure}
  \begin{subfigure}[b]{0.4\textwidth}
  \includegraphics[width=1\textwidth]{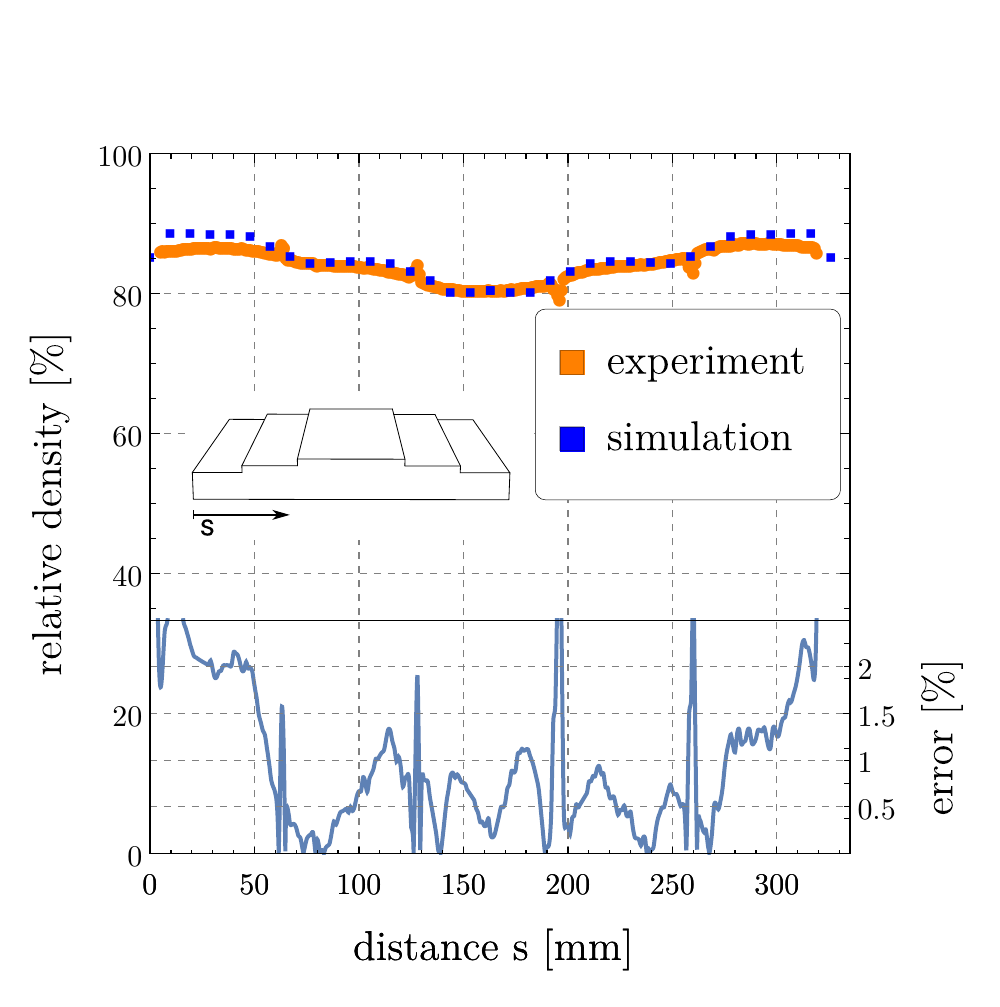}
  \caption{Density variation}
  \label{fig:TileDensity20}
  \end{subfigure}
  ~
  \begin{subfigure}[b]{0.4\textwidth}
  \includegraphics[width=1\textwidth]{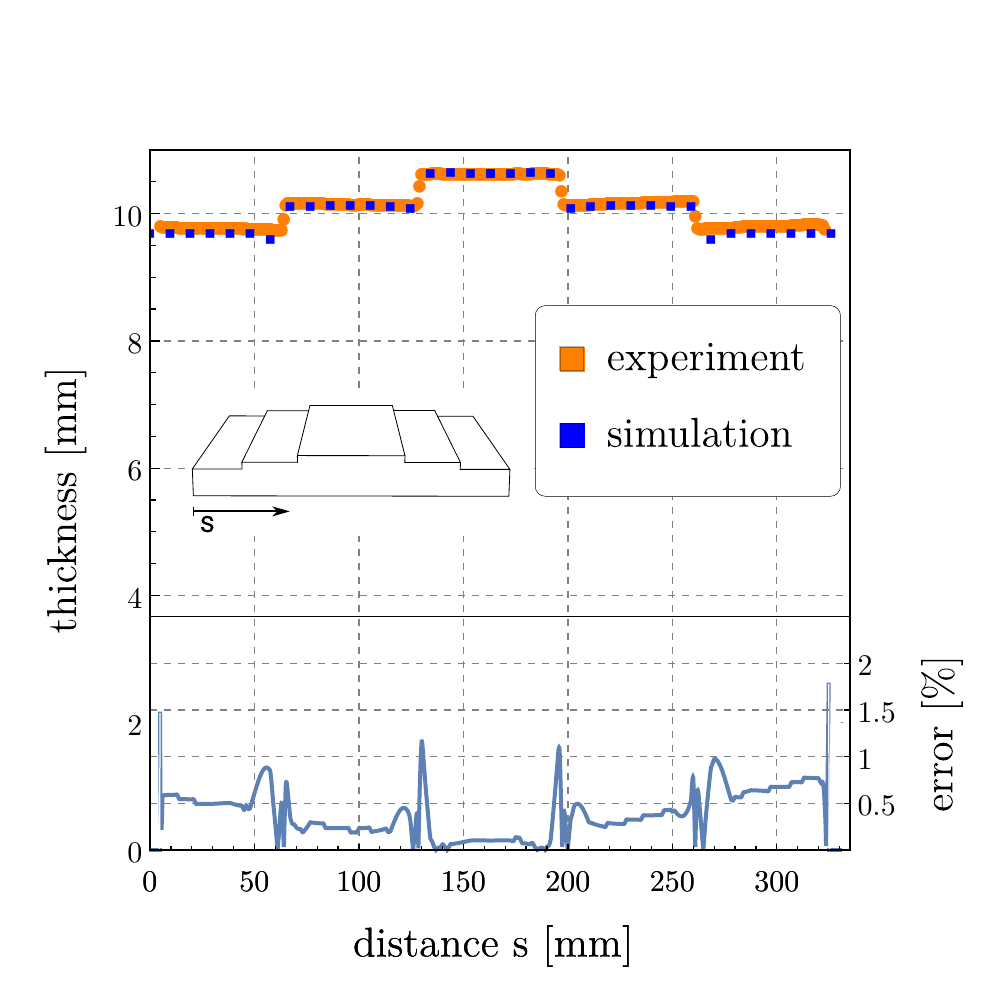}
  \caption{Thickness variation}
  \label{fig:TileThickness20}
  \end{subfigure}
  \caption{Density and thickness variation of the green after pressing, measured by an X-ray scanner, and compared to the model prediction through numerical simulation.
	Percent errors between results from the simulation and the experiment are reported in the lower parts of the graphs.}
 \label{fig:TileThickAndDens20}
 \end{figure}

\begin{figure}
  \begin{subfigure}[b]{0.4\textwidth}
  \includegraphics[width=1\textwidth]{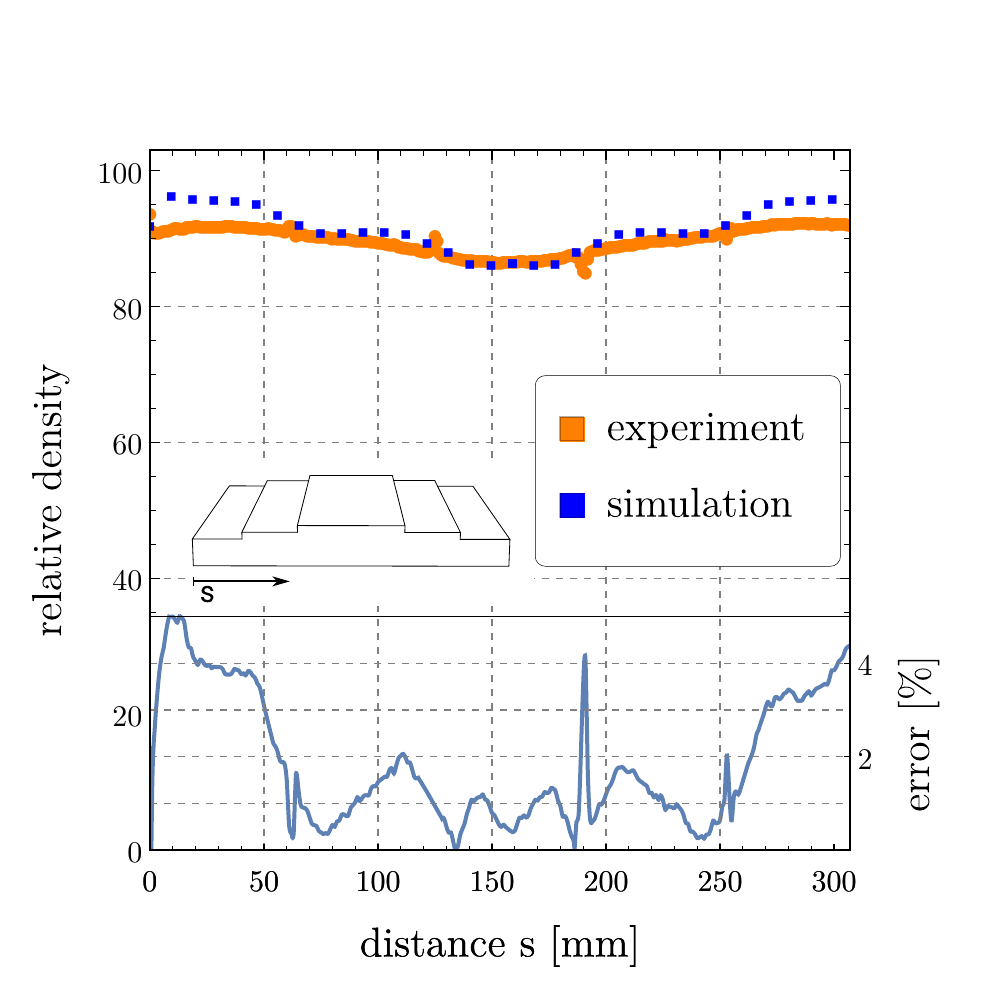}
  \end{subfigure}
  ~
  \begin{subfigure}[b]{0.4\textwidth}
  \includegraphics[width=1\textwidth]{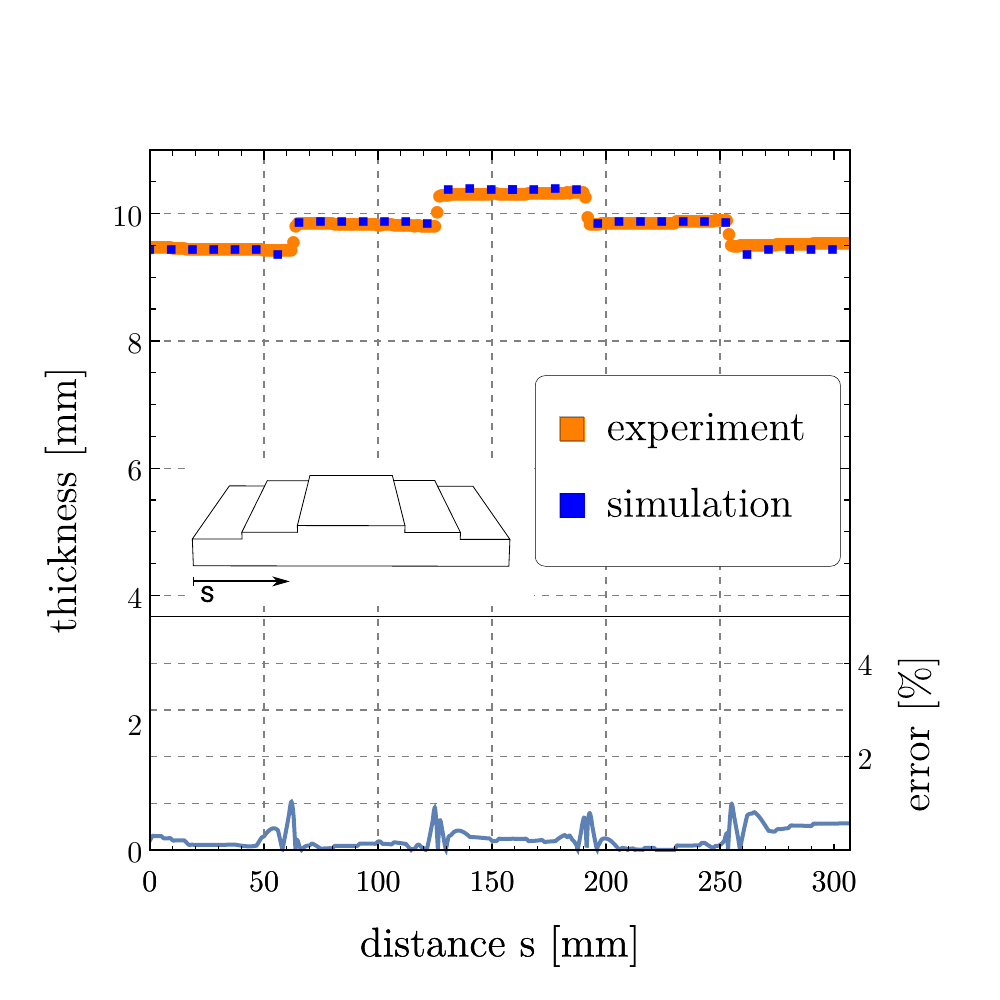}
  \end{subfigure}
\\
  \begin{subfigure}[b]{0.4\textwidth}
  \includegraphics[width=1\textwidth]{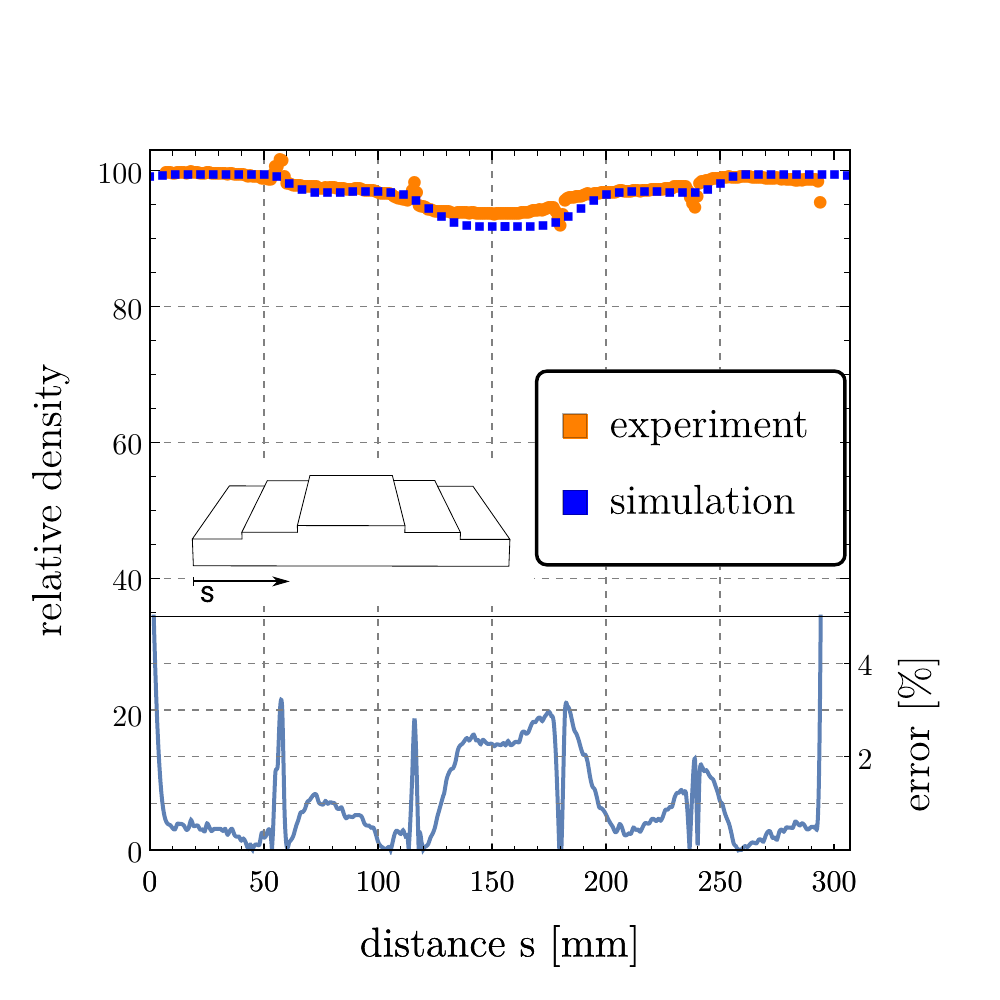}
  \end{subfigure}
  ~
  \begin{subfigure}[b]{0.4\textwidth}
  \includegraphics[width=1\textwidth]{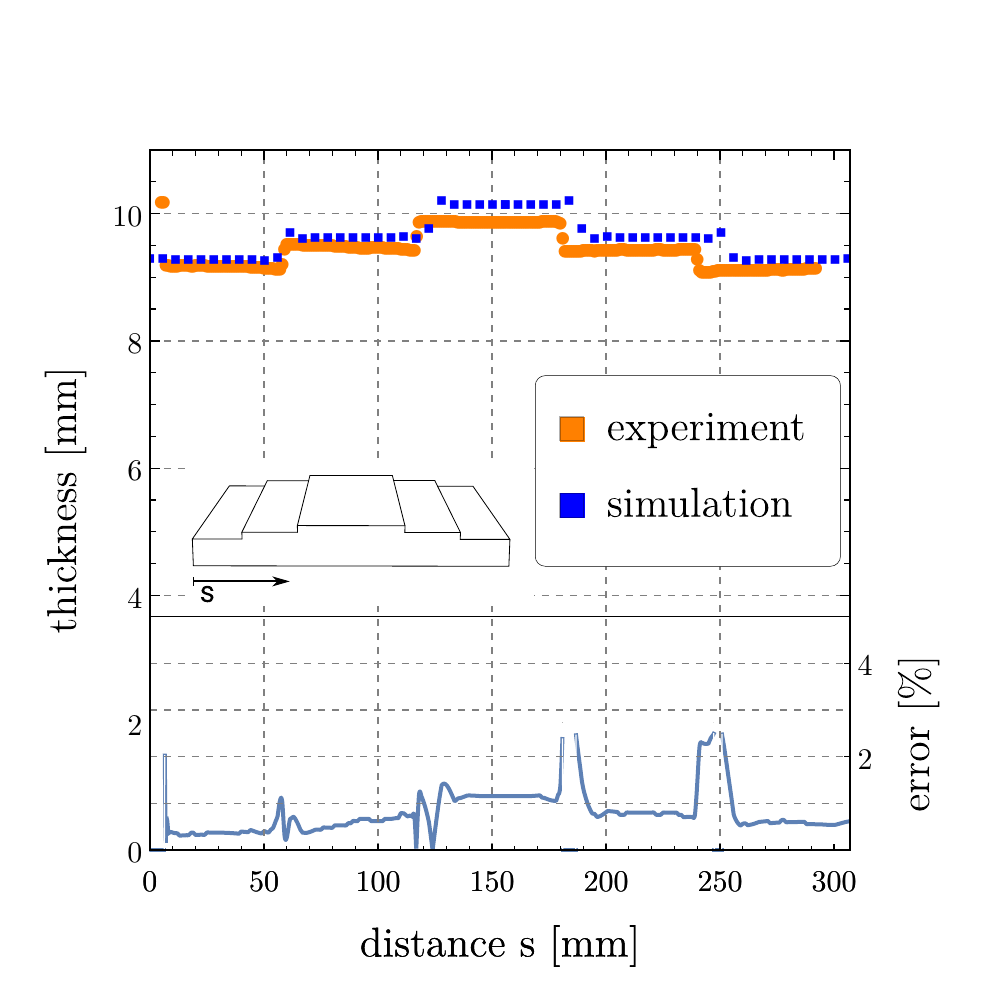}
  \end{subfigure}
\\
  \begin{subfigure}[b]{0.4\textwidth}
  \includegraphics[width=1\textwidth]{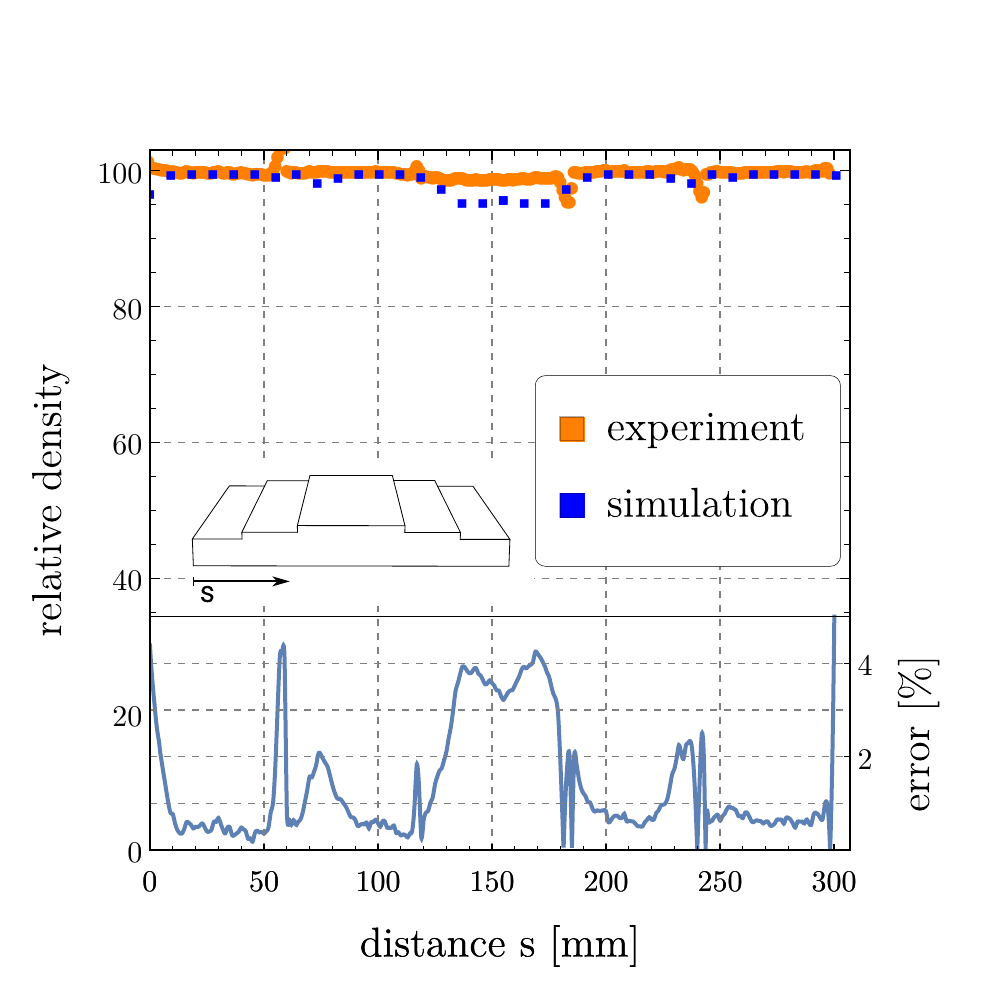}
    \caption{Density variation}
  \end{subfigure}
  ~
  \begin{subfigure}[b]{0.4\textwidth}
  \includegraphics[width=1\textwidth]{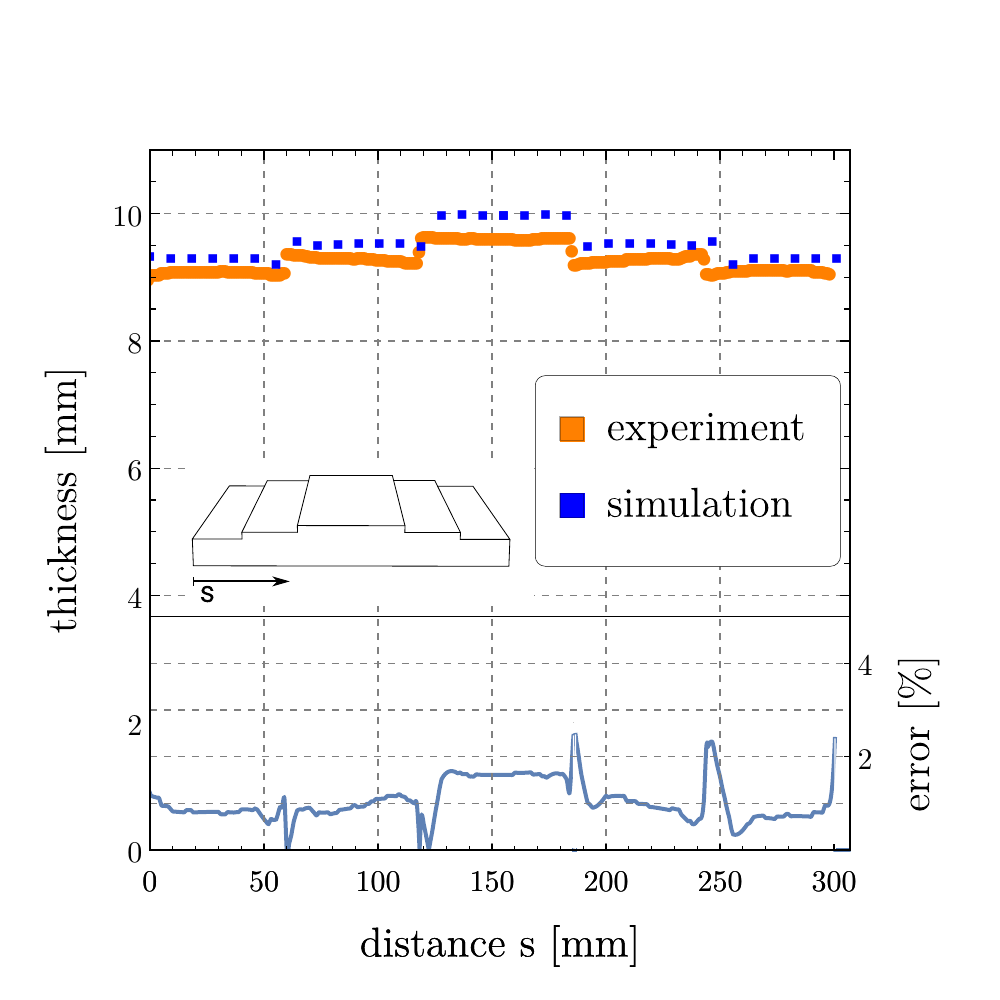}
    \caption{Thickness variation}
  \end{subfigure}
  \\
 \caption{Density and thickness variation of the ceramic piece after firing at 1100$^\circ$C (upper part), 1150$^\circ$C (central part) and 1200$^\circ$ C (lower part) measured by an X-ray scan, and compared to the model prediction through numerical simulation.
Percent errors between results from the simulation and the experiment are reported in the lower parts of the graphs.}
  \label{fig:TileThickAndDens1100-1200}
 \end{figure}

Measurements of density and thickness are shown and compared to the simulation in Figs. \ref{fig:TileThickAndDens20} and \ref{fig:TileThickAndDens1100-1200}. 
These figures prove the validity of the model, which is capable of reproducing the entire compaction and sintering processes of a ceramic granulate with an excellent precision, so that the percent errors 
(reported in the lower parts of the graphs) are  below 4\% at maximum and for most of the part below 2\% .

\begin{figure}[ht]
	\begin{center}
 \includegraphics[width=0.7\textwidth]{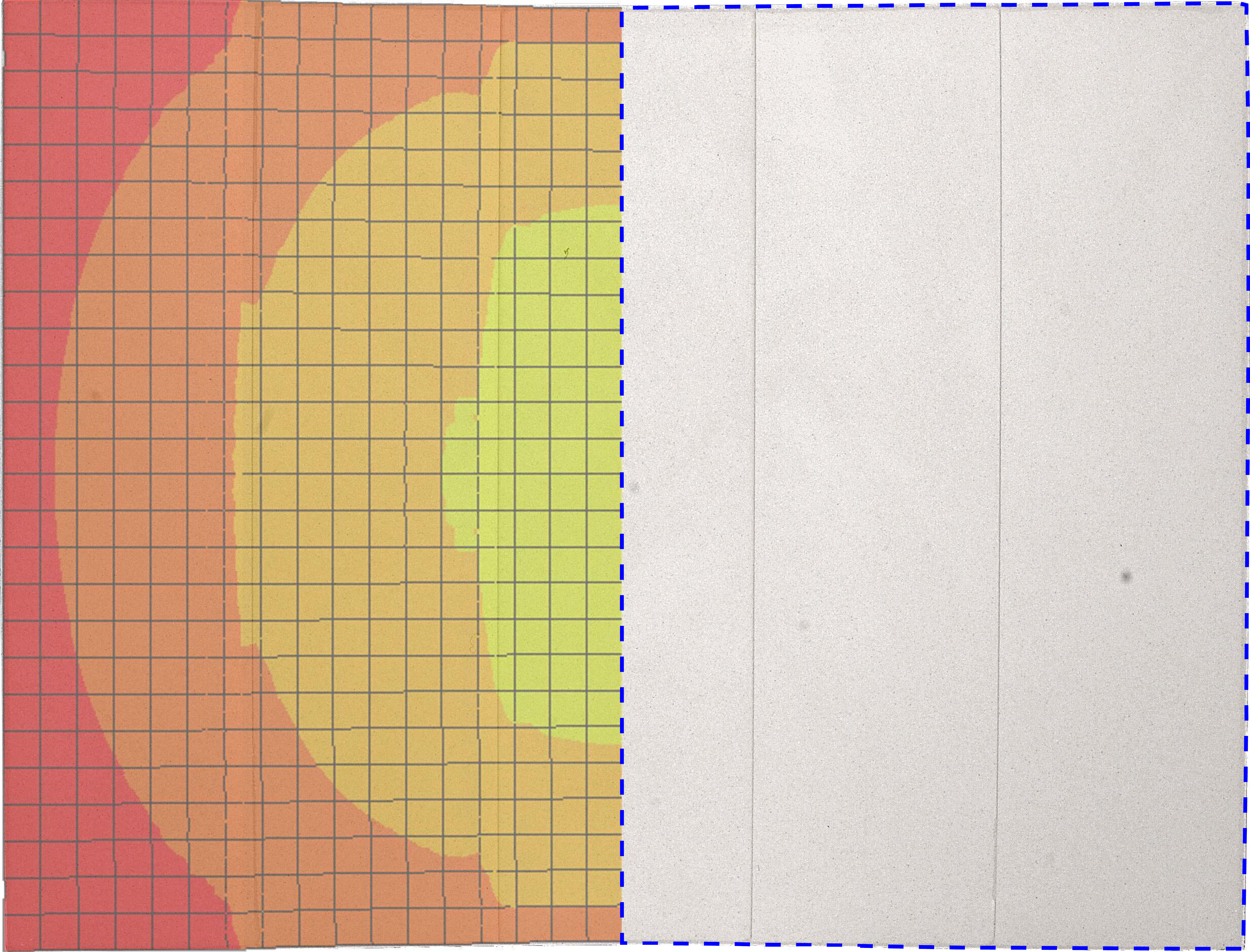}
\end{center}
\caption{The simulation of the formed and fired ceramic piece (on the left) compared to a photo of the real ceramic piece, sintered at 1200$^\circ$C (on the right, with the contour marked in blue). 
The qualitative trend of the distorsion of the boundary is well reproduced by the simulation, which also correctly captures also the pronounced shrinkage at the middle of the ceramic piece.}
 \label{fig:PhotoSinteredTile}
 \end{figure}

Finally, a photograph of the sintered ceramic piece is shown in Figure \ref{fig:PhotoSinteredTile} on the right, while the deformed mesh obtained after the simulated compaction  and sintering processis shown on the left. 
The contour of the photo shown in the figure has been marked in blue, so that it can be observed that the distorsion of the boundary induced by sintering is well captured by  the model, another demonstration of the excellent predictive capabilities of the mechanical model.

\clearpage

\section{Discussion and conclusions}

It has been shown that a thermomechanical model can be formulated, implemented and calibrated to provide a computational tool for the 
simulation of the entire process of ceramic production, starting from the cold pressing of a granulate and ending with the subsequent non-isothermal firing and sintering. 

Compared with the results of {\it ad hoc} performed experiments, the model predictions provide an accurate description of the density distribution 
and the shape distorsion suffered by the piece during the production process. 

Although the developed model is based on several simplificative assumptions and the lack of experimental data (inherent to the 
extreme conditions to which a ceramic piece is exposed) has precluded a fine calibration of model parameters, it is believed that the presented results 
show that mechanical modelling is a valuable alternative to the empirical processes still often in use for the design of pieces in the ceramic industry.

\paragraph{Acknowledgements.}
The authors would like to acknowledge invaluable assistance and guidance during experiments provided by Ing. 
Claudio Ricci (Sacmi, Imola) and Dr. Simone Sprio (CNR-Istec, Faenza). 
The authors are grateful to Professor Rebecca Brannon (University of Utah) for her kind advice, particularly on the treatment of viscoplastic constitutive equations. 

Financial support from FP7-PEOPLE-2013-ITN Marie Curie ITN transfer of knowledge programme PITN-GA-2013-606878-CERMAT2  (D.K.), from PRIN 2015 'Multi-scale mechanical models for the design and optimization of micro-structured smart materials and metamaterials' 2015LYYXA8-006 (D.B.) and from the ERC advanced grant ERC-2013-ADG-340561-INSTABILITIES 
(A.P.) is gratefully acknowledged.

\clearpage

\printbibliography 

\end{document}